\documentclass[twocolumn,showpacs,amsmath,amssymb,superscriptaddress,footinbib]{revtex4-1}

\usepackage{hyperref}

\usepackage{bibentry}

\usepackage{natbib}
\usepackage{color}

\usepackage{graphicx}
\usepackage{dcolumn}
\usepackage{bm}
\usepackage{enumerate}
\usepackage{slashed}
\usepackage{simplewick}
\usepackage{comment}

\usepackage{soul}

\usepackage[english]{babel}
\usepackage{appendix}
\usepackage[utf8]{inputenc}

\usepackage{marginnote}
\usepackage[normalem]{ulem}

\usepackage{subfigure}

\newcommand{\bz}{\bar z}

\newcommand{\ii}{{\rm{i}}}

\newcommand{\nn}{\nonumber}

\newcommand{\eq}[1]{(\ref{#1})}
\renewcommand{\>}{\rangle}
\newcommand{\la}{\label}
\newcommand{\ba}{\begin{align}}
\newcommand{\ee}{\end{equation}}
\newcommand{\be}{\begin{equation}}

\def\12{\frac{1}{2}}
\newcommand{\p}{\partial}
\newcommand{\en}{\end{align}}

\newcommand{\<}{\langle}

\usepackage{color}

\usepackage{amsmath}

\begin{document}

\setcounter{secnumdepth}{-1}

\title{Emergent Conformal Symmetry and transport properties of Quantum Hall States on Singular surfaces}
\author{T. Can}
\affiliation{Simons Center for Geometry and Physics, Stony Brook University, Stony Brook, NY 11794, USA}

\author{Y. H. Chiu}
\affiliation{Kadanoff Center for Theoretical Physics, University of Chicago, 5640 South Ellis Ave, Chicago, IL 60637, USA}
\author{M. Laskin}
\affiliation{Kadanoff Center for Theoretical Physics, University of Chicago, 5640 South Ellis Ave, Chicago, IL 60637, USA}
\author{P. Wiegmann}
 \affiliation{Kadanoff Center for Theoretical Physics, University of Chicago, 5640 South Ellis Ave, Chicago, IL 60637, USA}

\date{\today}

\begin{abstract}

We study quantum Hall states on surfaces with conical singularities.
We show that a small parcel of electronic fluid at the cone tip gyrates with
an intrinsic angular momentum whose value   is quantized in units of the
Planck constant  and exists solely due the gravitational anomaly. We show
that quantum Hall states  behave as conformal primaries near singular points,
with a conformal dimension equal to the angular momentum. We argue that the
gravitational anomaly and conformal dimension   determine the fine structure
of electronic density at the tip. The singularities emerge as quasi-particles
with      spin and  exchange statistics  arising from adiabatically braiding
conical singularities. Thus, the gravitational anomaly, which appears as
a finite-size correction on smooth surfaces, dominates geometric transport
on singular surfaces.

\end{abstract}

\date{\today}

\maketitle


\newpage
\noindent{\bf Introduction} Early developments in the  theory of the quantum Hall effect
\cite{Seiler1995,Levay,*Levay1997,Wen1992a,Frohlich1992} as well as a recent resurgence \cite{HoyosSon,Read2009,*Read2011,Douglas2010,*Klevtsov2013,CLW,CLWBig,Abanov2014,*GromovAbanov,*framinganomaly,lcw,Klevtsov2014,Read2015,KW2015} point to the role of geometric response as a fundamental probe of quantum liquids with topological characterization, complementary to the more familiar electromagnetic response. Such liquids exhibit    non-dissipative transport as a response to variations of the spatial geometry, controlled by quantized transport coefficients. This {\it geometric transport} is distinct from the transport caused by electromotive forces.
It is determined by the  geometric transport coefficients which are  independent characteristics
of  the state  and can not be read  from the electromagnetic response. 

 Surfaces with a singular geometry, such as isolated conical singularities,  or disclination defects, highlight the  geometric properties of the state. For this reason, they serve as an ideal setting to probe the geometry of QH states.  In this paper, we demonstrate this by examining Laughlin states on a singular surface, where geometric transport is best understood. We compare     spatial curvature singularities
 to magnetic ones (flux tubes) and   emphasize the difference. While the QH state imbues both types of singularities with local structure such as charge, spin, and statistics,  {\it only}  the curvature singularities reflect the geometric transport. 

 The gravitational anomaly is central to understanding the geometry of topological states \cite{CLW,CLWBig,GromovAbanov,lcw,Klevtsov2014,Read2015,KW2015,Vafa}. This
effect encodes the geometric characterization of such states,
 and is often referred to as the {\it central charge}.  

On a smooth  surface  the gravitational anomaly is a sub-leading effect. For
example, the central charge, $c_H$, appears as a finite size
correction to the angular momentum of the electronic fluid. 
 In \cite{KW2015} it is shown that the  angular momentum on a small region $D$ of the fluid on a surface with rotational symmetry is 
\begin{align}
 \mathrm{L} =  -\frac {1}{2\pi}\int_{D}\left({\mu_H}  (eB)-  \hbar\frac{c_H}{24}
{R}\right)dV,\la{1}
\end{align} 
 The last term in \eq{1} represents the gravitational anomaly. We are ultimately interested in features which do not depend on the size of the domain $D$. For the $j$-spin Laughlin states (see \cite{CLW}, and \eq{12} below for the definition of spin) the transport coefficient $\mu_H$ and the `central charge'  \(c_H\)  were found to be\begin{align}
c_H = 1 - \frac{12\mu_H^2}{\nu},\quad  \mu_H = \frac 1 2 (1 - 2j\nu),\la{c}
\end{align}  
where \(\nu\) is the filling fraction. 

 On a smooth surface, the geometric transport is hard to detect, since $c_H$ typically enters transport formulas as a small correction. We want to identify a setting where the gravitational anomaly $c_{H}$ is the  dominant feature,  as opposed to being a finite-size correction
overshadowed by a larger electromagnetic contribution.  We demonstrate that a surface with conical singularities  brings geometric transport to the fore.

We  show that a   small parcel of electronic  liquid centered at a singularity
spins  around it with an {\it\ intensive} angular momentum.  It is unchanged
when the parcel   volume $D$ shrinks as long as it stays larger than the
magnetic length. The intensive angular momentum is universal and  proportional
to \(c_H\), the geometric characteristic of the state  (see (\ref{9},\ref{171})). Similarly,  the  parcel maintains the excess or deficit of electric
charge and moment of inertia in the limit of vanishing $D$. That is to say, the moments are  localized near the conical point, see (\ref{1311},\ref{11}).    
Besides the charge, this property leads to the notions of spin and exchange statistics
of singularities. We compute both, see (\ref{10},\ref{1234}), and show that they are also determined
by the gravitational anomaly.

 Our argument stems from the observation that a state near the singularity has conformal symmetry.  Specifically, we find that it transforms as a primary field. 
 
 {Singularities elucidate the uneasy relation of QH-states to
 conformal field theory.  In general, QH-states do not possess conformal symmetry.
  They feature   
a scale - the magnetic length. As a result, physical observables do not transform conformally. 
However, the states appear
to be conformal in the vicinity of a singularity. In this paper, we show that the state is primary with a conformal dimension identical to the  angular momentum in units  $\hbar$. We emphasize that the dimensions have 
the {\it\ opposite} sign to the dimensions of similar primary fields in conformal field theory

 Conical singularities are not as exotic as they may seem, and occur naturally in several experimental settings. Disclination defects in a regular lattice can be described by metrics with conical singularities \cite{Katanaev1992}, and occur generically in graphene  \cite{graphene1,*graphene2}. In a recent photonic experiment,  synthetic Landau levels on a cone were designed in an optical resonator \cite{Simon2015}.

 A conical singularity of order \(\alpha<1\) is an isolated point $\xi_0$ on the surface with a
concentration of curvature\begin{align}\la{R}
R(\xi) = R_0 +4 \pi \alpha \delta (\xi - \xi_0),
\end{align}
 where $R_0 $ is the background curvature, a smooth function describing the curvature away from the singularity and the delta-function is defined such that its integral over  the volume element gives one. We refer $\xi_{0}$ as the conical point.  
 
%
%
 Examples of  genus-zero surfaces with constant curvature and conical singularities include: $R_0
> 0$  - an `american football' with two antipodal conical 
singularities
\cite{Troyanov1989}, \(R_0 = 0 \) - a  polyhedron  \cite{Troyanov2007,
Thurston1998}, \(R_0 < 0\) - a pseudo-sphere  (see e.g. \cite{TZ2002} and references therein).   
For the purpose of this paper it suffices  to consider conical singularities
which locally are flat surfaces of revolution.  If 
 \(\alpha >0\) the  singularity is equivalent to  an embedded cone with the
apex angle  $2\arcsin\gamma$, where \(2\pi\gamma=2\pi(1-\alpha)\)
is called cone angle (see Fig.\ref{Fig1}). If \(\alpha<0\),
the singularity  can be seen as a branch point of   multi-sheeted Riemann
surface (though non-convex polyhedra also can contain cone points with degree
$\alpha <0$). 

An especially interesting case occurs when $\gamma$  or
$1/\gamma$ is an integer. In this case, it may be possible to represent the surface as an orbifold, a surface quotiented by a discrete group of automorphisms. Then the conical singularities arise as fixed points of the group action \cite{Thurston1998}. 

Most of the formulas below are valid regardless of whether concentrated curvature is positive or negative (given by the sign of $\alpha$),  but strictly do not apply for cusp singularities for which $\alpha = 1$. Braiding of singularities on orbifolds is more involved (see \cite{Knizhnik1987,
 Bershadsky1988} for a similar issue in the context of CFT). We do not address them here.

  Conical singularities affect QH states differently than magnetic singularities  (flux tubes)
 \begin{align}\label{B}
 eB(\xi)=eB_0 -  2\pi \hbar \, a  \delta (\xi - \xi_0),\quad 
 \end{align}
%
%
%
%
%
%
To emphasize the difference  between geometric and magnetic singularities we   consider both simultaneously: a magnetic flux
\(a\) threaded through the  conical singularity \(\alpha\).
 We take \(eB_{0}\) to be  positive throughout the
paper.

Lastly,  we comment on the inclusion of spin $j$. As  discussed in \cite{CLWBig,lcw,KW2015}    Laughlin states are characterized
 not only by the filling fraction but also by the spin. Spin
does not enter electromagnetic transport. {Nor does it}
enter local bulk correlation functions, such as the  structure factor.
The spin enters  the geometric transport as {seen in} \eq{c}. 

To the best of our knowledge, there is no  experimental or numerical 
evidence that determines the spin in QH materials,  
nor are there any arguments that $j=0$, as it silently assumed in earlier papers. {For this reason, we keep spin as
a  parameter. It affects the physics of the QHE.  For example, at the filling \(\nu=1/3\),} the central charge  vanishes at \(j=1,\text{and}\,j=2\).  The central charge  equals $-2$ if $\nu=1$ and $j=0$ or $1$. If \(j=\frac 1{2\nu}\), the coefficient  \(\mu_H\) vanishes and \(c_H=1\).
 \smallskip

\noindent{\bf Main results}   
{\it a.\,Conformal dimensions}. 
 We show that the states are  conformal primary in the vicinity of a singularity, magnetic or geometric.
In  \cite{CLW,lcw}  (see also \cite{Kvorning}) it was shown that the magnetic singularity is a conformal primary with the dimension 
\begin{align}
h_a= \frac 12 a(2\mu_H-\nu a).\la{8}
\end{align}
  In this paper we show that the geometric singularity is  also
conformal primary, but in this case its dimension is controlled solely  the gravitational
anomaly
\begin{align}
\Delta_\alpha= \frac {c_H}{24}(\gamma^{-1}-\gamma),\quad \gamma=1-\alpha.\la{9}
\end{align} The formula \eq{9} is familiar in the conformal field theory: \(-\Delta_\alpha\)    (mind the opposite sign!) is the  dimension of a  vertex operator  {of} a conical point in conformal field theory  with the central charge $c_H$ \cite{Knizhnik1987,Bershadsky1988}. The same formula
 enters the finite size correction to the free energy of critical systems
on a conical surface  \cite{Cardy1988} and equivalently the formula for the determinant of the Laplace operator (e.g., \cite{Kokotov,Aurell2}). These
are not coincidences.  In the neighborhood of a  singularity, QH-states and conformal field theory share the same mathematics, but are by no means identical: the conformal dimension of QH states is opposite to that in conformal field theory with the central charge given by \eq{c}.

  Conformal dimensions are    important characteristics which enter  physical
observables. 
\smallskip

{\it b.\, Dimension, gyration, and spin}
 We show that the dimension determines transport in the neighborhood of the singularity.  The electronic fluid gyrates around the conical point with an intensive angular momentum, independent of volume. 
 We will show that   the intensive part of the angular momentum   is  exactly the dimension  \eq{9}\begin{align}
 \mathrm {L_\alpha}=\hbar\Delta_\alpha. \la{171}
\end{align} 
 A reason for this is that  the angular momentum of the gyrating fluid gives  the spin to the singularity - 
an adiabatic rotation of the state by \(2\pi\) results in
a phase \((2\pi/\hbar) L_{a,\alpha}\).  But because the state is holomorphic, its spin is identical to the dimension.

A similar formula holds  for the angular  momentum of the combined magnetic and geometric  singularities \begin{equation} \mathrm {L}_{\alpha,a}=\hbar\left(\frac
 1\gamma h_a+\Delta_\alpha\right).\la{172}\end{equation}

The intensive angular momentum \eq{172}  is added to
\eq{1}
Hence, if the area $D$ of the fluid parcel is taken to zero, only the angular momentum  \eq{172} remains.

\smallskip

{\it c.\,Braiding singularities} Just like  Laughlin's quasi-holes (which are closely related to flux tubes \eq{B}), conical singularities can be braided. The phase acquired by adiabatically exchanging two singularities is called the exchange  statistics. Braiding two quasi-holes with charges \(a_1\) and  \(a_2\)
yields the phase  \be\Phi_{12}=\pi (\nu a_1a_2).\ee This result is known since
early days of QHE  \cite{Arovas1984}. 

Braiding conical  singularities is more involved. We argue
that   braiding phase  of two cones of the order \(\alpha_1\) and \(\alpha_2\) 
{are determined  exclusively by the central charge} \begin{align}\nn
\Phi_{12}=-\pi & \frac{c_{H}}{24} \alpha_{1}\alpha_{2}
\left( \frac{1}{\gamma_{1}}
+ \frac{1}{\gamma_{2}} \right)=\\&\pi\left(\alpha_{2} \Delta_{\alpha_{1}} + \alpha_{1} \Delta_{\alpha_{2}}\right)
+\pi \frac{c_{H}}{12}  \alpha_{1} \alpha_{2.} \la{10}
\end{align} 
Here, we assume that the path is sufficiently small, so conical singularities are the
only contributions to the  solid
angle swept out by the path. The first two terms in \eq{10}  are  the phase acquired by a particle with  spin
 \(\Delta_{\alpha_1}\) (or \(\Delta_{\alpha_2}\)) going half way around a solid angle \(4\pi\alpha_1\)(or \(\alpha_2\)). The last term 
 \[\frac{c_{H}}{12}  \alpha_{1} \alpha_{2}\]
  is the exchange statistics.  
  
  On an orbifold, where { either} \(\gamma\) { or} \(1/\gamma\) is an integer \(n\)
the phase  for identical cones is  \be\Phi_{12}=\pi \frac{c_H}{12}\left(\sqrt n-\frac 1{\sqrt n}\right)^2.\la{1234}\ee
It appears  rational,  even
in the case of the integer QHE.
\smallskip

The formulae (\ref{9}-\ref{10}) are our main results: the braiding statistics
of the singularities and the angular momentum of the electronic fluid around  a cone are {given solely by} the gravitational anomaly.
Other results such as the transport and the fine structure of the density
profile at the singularity are shown below.
\smallskip

{\it d.\,Moment of inertia}    The conformal dimension  can be also  read-off from the fine structure of the density profile in the neighborhood of the singularity. On a singular surface  the density changes abruptly on the scale of magnetic length and in the limit of 
 vanishing magnetic length   is  a singular function. {It} is  properly characterized
 by the moments  
\begin{align}
m_{2n} = \int (r^2/2l^2)^{n} (\rho - \rho_\infty) dV.\la{81}
\end{align}Here,  \(\rho_\infty=\nu
 (e/h)B\) is 
 the asymptotic value of the density away from the singularity and \(l=\sqrt{\hbar/(eB)}\) is the magnetic length. In the integral  \eq{81}  $r$ is the Euclidean  distance to
the singularity and  $dV = 2\pi\gamma rdr$ is the volume element. 

The first moment, the `charge' \(m_0\),    follows from
 the generalized St\v{r}eda  formula -- the number of particles
in an area $dV$  is saturated by $\bar{\rho} dV$ where
\begin{align}\la{111}
 \bar \rho =\nu( eB/h) + (\mu_H/4\pi) R.
\end{align}
 We will obtain this relation in the next section. 

Hence \begin{align}
m_{0} =\int { ( \bar\rho - \rho_{\infty}) }\,dV=-\nu a+  \mu_{H} \alpha.\la{1311}
\end{align}
   Eq. \eq{1311} says that if \(\mu_H\!>\!0\), the apex accumulates electrons  when $\alpha >0$.   It  gives an alternative
definition of the transport coefficient   \(\mu_H\).This result for  \(j=0\)
 is well  known (see, e.g.,  \cite{Biswas2014,Wiegmann2013,Abanov2014}) and there is even a recent claim  
 of  experimental observation \cite{Simon2015}. However, the gravitational anomaly does not enter here. It emerges in the
 next   moment, the {\it  moment of inertia}  of the gyrating parcel \(m_2\).
We  will see that  \begin{align}
 m_{2}=(1-j)m_0+\gamma^{-1} h_a+\Delta_\alpha,\la{11}
\end{align}
where \(h_a \) and  \(\Delta_\alpha\) are the dimension  (\ref{8},\ref{9}). We check this formula against the 
 integer QH effect,  \(\nu=1\), where   all the moments are computed exactly. We
do this in the last section of the  paper.

The relation between the moment of inertia \eq{11} and the angular moment
 \eq{171} is not surprising.  In a QH state,  positions of  particles determine their velocities. Consequently,
the density determines the momentum \(\mathrm{P}\) of the flow. The  relation between the momentum and the density has been  obtained in \cite{Wiegmann2013,*PW12,*PW13},  in  the next section we recall its origin. 
The relation  reads\begin{align}
\nabla\times \mathrm{P}=-eB \left(\rho-\bar\rho\right)+\frac \hbar2 (1-j)\Delta\rho,\la{151}
\end{align}
 where  \((\nabla\times)_i=\epsilon_{ij}\nabla_j\), {} \(\nabla_j\)
is a covariant derivative,  \(\Delta\) is the Laplace-Beltrami operator,
and \(\bar\rho \) is given by \eq{111}.    In the next section we recall its origin. 

With the help of this formula we express the angular momentum in terms of the density. In order to avoid unnecessary complications with the definition
of the angular momentum on a curved surface, we assume that close to the singularity the surface locally can be approximated by a flat  surface with rotational symmetry.  Then the  angular momentum 
about the conical point, expressed in local coordinates \(\xi\), is given by the standard formula
\(\mathrm{L}=\int(\xi\times\mathrm{P})\,dV\).

Using \eq{151} we obtain 
 \begin{align}\mathrm{L}\!=
\!  (eB)\int \frac{r^2} 2(\rho-\bar\rho) dV+\hbar(j-1)\int \rho dV.\la{150}
\end{align} 
 Interpreting   this formula we notice that  the first term  is  the diamagnetic
effect  of 
fluid gyrating in magnetic field the second term is the paramagnetic contribution.

 The formula for the charge of the cone  \eq{1311} is
a  consequence of \eq{151}. Away from the singularity the momentum
rapidly vanishes. As a result the integral  \(\int (\nabla\times \mathrm{P}) \,dV\) vanishes. Then \eq{151} yields  \eq{1311}.

The integral \eq{150}  over the
bulk of the surface gives the  extensive part \eq{1}, while the integral
over a patch at the singularity is  \(m_2\!-\!(1\!-\! j)m_0\). Then   \eq{11} yields  \eq{172}. It remains to compute (\ref{8},\ref{9}).

\smallskip

{\it e.\,Transport at the singularity}.  Since the work of Laughlin \cite{Laughlin1981}
it
was known
that an  adiabatic  change of the magnetic flux \(a(t)\) in \eq{B} threading
through the puncture
of a disk causes a  radial electric current {flowing outward}    \(\mathrm{I}=-\nu
e \dot
a\).   

Adiabatically evolving the order of the conical singularity \(\alpha(t)\) 
{also induces} a current.
It follows from   \eq{1311} {that the current flowing away from} the apex
\(\mathrm{I}=e\dot  m_0\) is   \(\mathrm{I}=e\mu_H\dot\alpha\).    

More interestingly, both evolving flux and the cone angle accelerate the
gyration of the fluid, and produce a torque. The torque is the  moment of the force  exerted on  a  fluid parcel
    \(\mathrm{
M}=\int
(\mathrm{r}\times \mathrm{F})dV\). Since \(\dot{\mathrm P}=\mathrm{F}\),  the torque
is  the rate of change  of the angular momentum  \(\mathrm{M}=\dot{\mathrm L}\).  From  \eq{171} it then follows that the torque is proportional to the rate of change of the conformal  dimension.
We collect
the formulae for electric and  geometric transport
 \begin{align*}
 \text{e-transport:}\;\; &\text{current}=-e\nu \dot
a,&&
\text{torque}=\hbar \dot h_a, \\  \text{g-transport:} \;\;&\text{current}= e\mu_H\dot\alpha,&&
\text{torque}=\hbar\dot\Delta_\alpha.
\end{align*}   
These formulas put geometric transport in a nutshell.
  
  \smallskip
  
In the remaining part of the  paper we obtain the dimensions (\ref{8},\ref{9}) and the statistics \eq{10} by  employing the  conformal Ward identity, a framework developed in {\cite{Zabrodin2006,CLW}. \smallskip

\noindent{\bf QH-states on a Riemann surface } Before
turning to  singular surfaces, we recall some key facts about
   Laughlin  states on a Riemann surface
\cite{Klevtsov2013,CLW}. 

  The most compact form of the state appears in 
locally chosen complex
coordinates \((z,\bar z)\), where the metric is conformal \(ds^2=e^{\phi}|dz|^2\).
 In these coordinates the  
  Laplace-Beltrami operator is \(\Delta=4e^{-\phi}\p_z\p_{\bar z}\) and
   the volume form is \(dV=e^{\phi}d^2z\).
 

In the conformal metric we can always choose coordinates such that the unnormalized spin-$j$ state   reads
\begin{align}\la{12}
\Psi \!=\!  \prod_{1\leq i<k}^N\!(z_i\-\!-\!z_k)^\beta \exp{
\sum_{i=1}^N  \frac 12\left[  Q(z_i,\bar z_i)\!\!-\!\!j\phi(z_i,\bar z_i)\right]}
\end{align} t
where, the integer $\beta=\nu^{-1}$ is the  inverse filling fraction and
\(Q \) is the magnetic potential defined by \(-\hbar\Delta Q=2eB\).

{\bf}  While the wave function \eq{12} explicitly depends on
the choice of coordinates, the normalization factor
\begin{align}\la{14}
&\mathcal{Z} [Q,\phi]= \int  |\Psi|^2\prod_i \exp{
\left[\phi(z_i,\bar z_i)\right]}d^2z_i
\end{align} 
 does not. It is an invariant functional depending on  the geometry of the surface, and
in particular on the positions and orders of singularities.  

The functional encodes the correlations and the transport properties of
the state and for that reason is referred to as a generating functional. For example, a variation
of the generating functional over  the  magnetic
potential \(Q\) at a fixed conformal factor \(\phi\) is the particle density 
 \begin{align}\rho\,dV= \left(\frac{\delta\log
\mathcal{Z}}{\delta
Q }\right)\,d^2 z. \nn\end{align}

  In \cite{KW2015}, it was shown that 
 that a variation of the generating functional over \(\phi\) at a fixed volume and a fixed   gauged potential implies the  variational formula for the momentum of the fluid
  \begin{align}\la{2010}
&\mathrm{P}=\frac\hbar  2 \nabla\times\left(
\frac{\delta\log \mathcal{Z}}{\delta
\phi}\right).
\end{align} 
Then for surfaces of revolution the quantity
\begin{align}
\mathrm{L}=- \hbar\int\left(\frac{\delta\log
\mathcal{Z}}{\delta
\phi}\right)d^2z.
\la{201}\end{align} 
is interpreted as angular momentum. In (\ref{2010},\ref{201}) the variation is taken at a constant magnetic field and curvature.

With the help of these formulas  we can obtain the relations   (\ref{151},\ref{150}).  They  follow from the observation
that the magnetic potential and the conformal factor  
appear in (\ref{12},\ref{14}) on almost  equal footing, besides that under  a variation
over the conformal factor   magnetic potential  varies as
\(-\hbar\Delta\delta Q=2\delta\phi(eB)\). This  contributes to the diamagnetic part of the relation 
\eq{151}.

\smallskip
 
\noindent{\bf QH-state on a cone} A surface  has  a conical
singularity of order \(-\alpha\)\,(\(\alpha<1)\) if in the neighborhood
of the conical point \(z_0\) the conformal factor behaves as \begin{align}\phi\sim
-\alpha
\log|z-z_0|^2.\la{24}\end{align}Locally a cone is thought as a wedge  of a plane with
 the deficit angle \(2\pi\alpha\), whose sides are
isometrically  glued together (see the Fig. \ref{Fig1}).
\begin{figure}[htbp]
\begin{center}
\includegraphics[width=8.5cm]{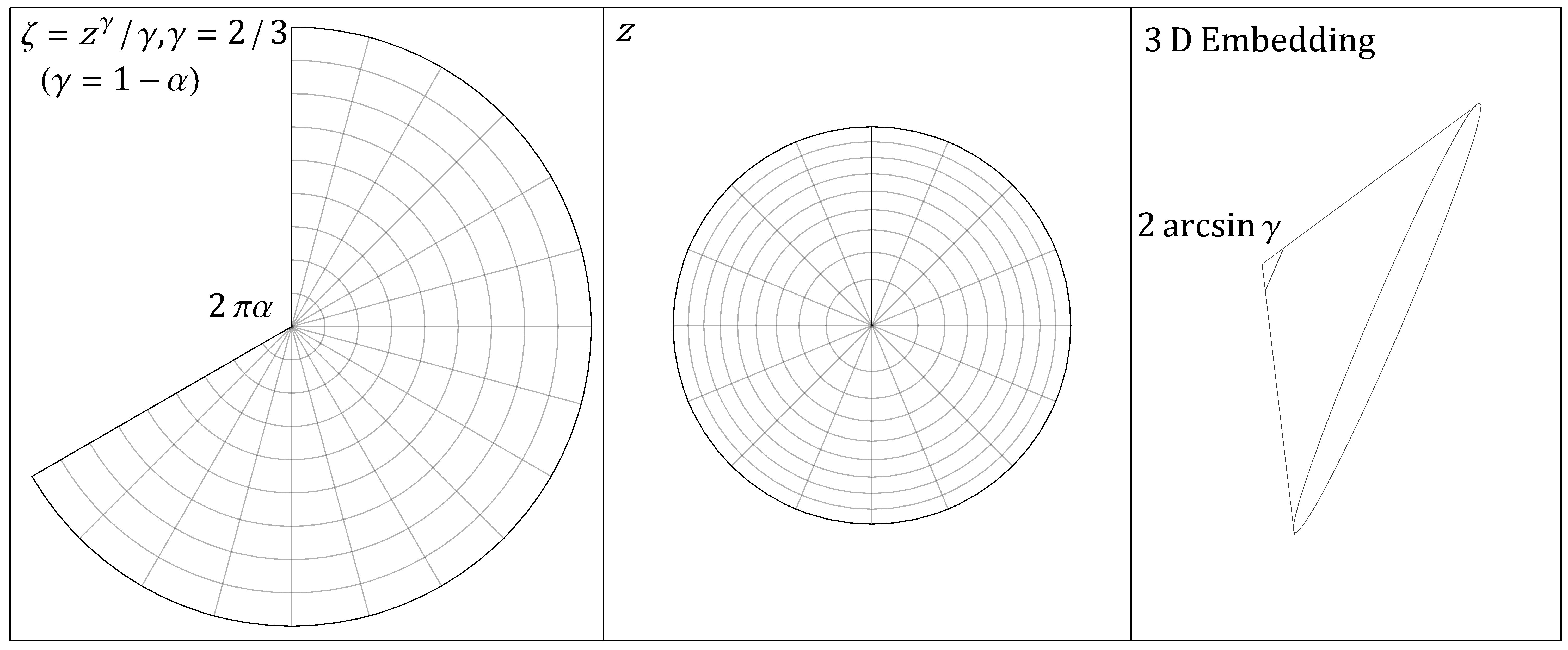}
\caption{Schematic diagram of a cone and its 3D embedding}
\label{Fig1}
\end{center}
\end{figure}
Let denote the complex  coordinate on the plane as \(\xi\) and the cone angle   \(2\pi\gamma=2\pi(1-\alpha)\). The wedge
is a domain \(0\leq{\rm arg}\, \xi<2\pi\gamma\) with the Euclidean metric
 \(ds^2=|d\xi|^2\). A pullback of a singular conformal
map \begin{align}z\to\xi(z) = (z-z_0)^{\gamma}/\gamma\la{131}\end{align}  maps the wedge to a punctured disk. The map
 introduces the complex coordinates \((z,\bar z) \) where  the  metric is conformal \begin{equation}ds^2=|z-z_0|^{-2\alpha}|dz|^2.\la{17}\end{equation} 
The quantum mechanics
on the cone assumes the `wedge-periodic' condition.  The lowest Landau level on a cone is
spanned by the holomorphic polynomials  of \(z\) (see, \eq{321})
in the metric \eq{17}.

Eq. \eq{12} is valid  on any genus-zero surface. Specifically, in
the neighborhood of the conical singularity  the  the conformal factor in \eq{12} behaves as \eq{24}   and locally the state reads
\begin{align}
\Psi_\alpha  \!=\!  \prod_{ i<k}\!(z_i\-\!-\!z_k)^\beta  \prod_{ i}\!|z_0\!-\!z_i|^{j\alpha}e^{-
 {|z_i-z_0|^{2\gamma}}/({4l^2\gamma^2})}.\nn
\end{align}
Then the  generating functional \(\mathcal{Z}_\alpha\)  is  the expectation value of this operator. 

A singularity can be interpreted as an insertion of the 
 `vertex
operator'  such that 
 the  generating functional \(\mathcal{Z}_\alpha\)  is  the expectation value of this operator.
 We will show that this operator is conformal primary.  This means that under a  dilatation transformation  of the  metric close to the singularity,
the functional transforms conformally   $$-\delta\log\mathcal{Z}_\alpha=\Delta_\alpha\delta\phi.$$ Here 
 \(\Delta_\alpha\) is the conformal dimension.
 Eq. \eq{201} identifies the  conformal dimension with  the
angular momentum  \eq{171}.  We compute it in the remaining part of the paper. 

 Calculations are the most convenient in complex notation.  We will need the formula for the angular momentum written in complex coordinates. The momentum  in complex coordinates   reads   \(P_{z}dz+
 P_{\bar{z}}d\bz=\ P_{\xi} d\xi + P_{\bar{\xi}} d \bar{\xi}\) in local coordinates related by the conformal map \eq{131}. Then with the help of \eq{131}, the angular momentum density in flat coordinates reads \( {\rm Im} (\xi P_{\xi}) = {\rm Im}(\xi \frac{dz}{d\xi} P_z)=\gamma^{-1} {\rm Im}(z P_z)\). Hence
 
\begin{align}
\mathrm{L}=-\gamma^{-1}\int {\rm Im}(zP_z)\,dV  \la{255}.
\end{align}

\smallskip

\noindent{\bf Conformal Ward identity} Moments of the density and  the angular
 momentum are computed via the Ward identity. The Ward identity reflects the invariance of the integral
\eq{14} under the infinitesimal  holomorphic change of variables $z_i\to z_i+\epsilon/(z-z_i)$. It
 claims that the  function of coordinates \(z_i \) and
a complex parameter \(z\) \begin{align*}
 \!\sum_{i} \frac{\partial_{z_i}Q+(1\!-\!j)\p_{z_i}\phi}{z - z_{i}} \!+\frac\beta
 2 \left( \sum_{i} \frac{1}{z - z_{i}}\right)^{2} \!+\!  \sum_{i} \frac{1-\frac\beta
 2}{(z - z_{i})^{2}}
\end{align*}
vanishes
 under averaging over the state.
 
The identity is closely related to  the Ward identity  of  conformal field theory. In order to   flesh out this analogy, we
introduce the  scalar field  \(\varphi\) 
\begin{align}& \varphi=-2\beta\sum_i\log|z-z_i| - Q.\la{241}
\end{align}
 We also need the holomorphic component of the  conformal  `stress tensor'  
\begin{align}T=\frac\nu
2 \<\left(\partial_z \varphi\right)^2\>  -\mu_H  \<\partial_z^2 \varphi\>.
\la{Tgrav}\end{align} 
 Then the Ward identity can be brought to the  form connecting the momentum and the  conformal  `stress tensor' 
 \begin{align}
\frac 1\hbar\int   \frac{\ii P_{z'}- \frac{\mu_H}{2\pi}  \p_{z'}(eB)}{z-z'}dV_{z'}=T.\la{27}
\end{align}
 We describe the algebra elsewhere.


\smallskip

\noindent{\bf Trace of the conformal stress tensor} The  meaning of the Ward identity 
 becomes transparent if we complete the  stress tensor by its trace \(\Theta\),
 defined  through the conservation law equation
  \begin{equation}\la{271}\p_{\bz}
T+e^{\phi }\p_z \Theta=0.\end{equation} 
Together the components \(T,\,\bar T,\,\Theta\) form a quadratic differential
\(T_{ij}dx^idx^j=T(dz)^2+\bar T(d\bz)^2+2e^\phi \Theta dz d\bz\).    
 Then  
\(\p_{\bz}\) derivative brings  \eq{27} to the form
 \begin{align}
 P_z=\frac  1{2 \pi \ii }\p_z\left(\mu_{H} (e B) -2\hbar \Theta\right).\la{29}
\end{align}
The formula \eq{29} identifies   the trace \(\Theta\)   with the intensive part
of the  angular momentum
 \begin{equation}{\mathrm{L}= -\frac{\mu_H }{2\pi\gamma} \int (eB )dV}+\frac\hbar{\pi\gamma} \int\Theta\,dV.\nn\end{equation}
\smallskip

\noindent {\bf Gravitational Anomaly} On its own, the Ward identity is a relation between one  and
 two-point correlation functions. The two-point function
\(\langle (\partial_z \varphi)^2 \rangle\) {in the Ward identity is evaluated coincident points}.
The connected part of this function \(T^A=\frac{\nu} 2\langle (\partial_z \varphi)^2 \rangle_c=\frac{\nu} 2\left[\langle (\partial_z \varphi)^2 \rangle-\langle \partial_z \varphi\>^2\right]\) is the {entry}
which converts the identity {into} a meaningful equation. In \cite{CLW} it
was argued that {the connected two-point function} is  proportional
to the Schwarzian   of the metric
 \begin{align}
T^A  =\frac 1{12} \mathcal{S}[\phi],\quad \mathcal{S}[\phi]\equiv-\frac 12 (\p_z\phi)^2 +\partial_z^2
\phi.\la{31}
\end{align}  
Thus \(T=T^C+T^A \) consists of the `classical' part\begin{align}
T^C=\frac\nu 2  \left(\<\partial_z \varphi\>\right)^{2}  -\mu_H  \partial_z^2 \<\varphi\> \la{32}
\end{align}and the  anomalous
part \eq{31}.  This explicit representation of $T$ converts the Ward
identity to the equation.

This equation consists of terms of a different order in magnetic length and
has to be solved iteratively.  The leading approximation, where \(\<\rho\>\approx\bar\rho\) suffices.   From \eq{241} it follows that  \(\<\varphi\>\approx (\mu_H/\nu)\phi\).  Up to this order   the classical part of the stress
tensor is \(
T^C= -(\mu_H^2/\nu)  \left[-\frac 12(\partial_z\phi)^2 +\partial_z^2
\phi\right]  \).
 Together with the anomalous  part  \eq{31} the stress tensor   reads\begin{align}
T=\frac{c_H}{12}  \mathcal{S}[\phi].\la{33}
\end{align} 
On smooth surfaces
by virtue of \eq{271}
\begin{align}
 \Theta=\frac{c_H}{48} R.\nn
\end{align} 
This  is the trace anomaly.

Thus,  to leading order the Ward identity is equivalent to the conformal
Ward identity.  Eq. \eq{29} then yields
 the result  obtained in \cite{KW2015} for the momentum on a smooth
surface
\begin{align}
 P_z=\frac 1 {2 \pi \ii} \p_z\left(\mu_H\, (e B) -\hbar   \frac{c_H}{24}
 R\right).\la{3}
\end{align}
In its turn it yields the angular momentum  given by \eq{1}.
For {reference} we present the formula for the density on a smooth surface, which follows from \eq{3} and \eq{151}. It {  was} previously obtained in
\cite{CLWBig}
\begin{align}\nn
\rho=\! \bar \rho \!+\!\frac 1{4\pi B}\left[ ( \nu\!-\!\frac 12)\Delta B\!+\!   \left ( (1\!-\!j)\frac{\mu_H}{2}\!+\! \frac{c_H}{24} \right) \frac \hbar e  \!\Delta R \right].  
\end{align}
 \smallskip
 
\noindent{\bf Geometric singularity} On a singular surface the formula  \eq{3} {is not valid, since the curvature is singular}. However, the results
promptly follow from the dilatation
sum rule  we now obtain.   

We multiply
  \eq{27} by \(\frac{
zdz}{2\pi \ii}\) and integrate it 
along a boundary of an infinitesemaly   small parcel.  Using \( \frac 1{2\pi\ii}\oint\frac{zdz}{z-z'}=
z'\), \eq{255}  we reduce   the  Ward identity to the sum rule for the intensive part of the angular momentum
\begin{align}
\gamma \mathrm {L}_{\alpha,a}=\hbar\,\oint T(z)\frac{
 zdz}{2\pi \ii}.\la{36}
\end{align}
and notice that in the neighborhood of  singularity  \(T(z)\) is a holomorphic function. Thus
 \(\oint T(z)\frac{
 zdz}{2\pi \ii}={\rm res} (z T)\). 

We compute the singular part of the
stress tensor by evaluating the Schwarz derivative on   the singular metric \eq{24}. 
Equivalently, we treat a  conical singularity   as a conformal map \eq{131}
and compute {the}  Schwarz derivative of the map   
 \begin{align}
\mathcal{S}[\phi]\equiv\{\xi,\,z\}=\frac{\xi'''}{\xi'}-\frac 32 \left(\frac{\xi''}{\xi'}\right)^2=\frac{\alpha(2-\alpha )}  {2z^2}.\nn
\end{align}

We obtain  \begin{align}
 T=\frac{c_H}{24}  \frac{\alpha(2-\alpha )}  {z^2}.{\la{256}}
\end{align}
Using \eq{36}, we arrive at our main result \eq{171}.
}
\smallskip

\noindent{\bf Magnetic singularity}   In this case, the gravitational anomaly does not contribute to the   the singularity of the the stress tensor. Rather, the stress tensor receives an additional contribution  from the magnetic potential of the flux tube $Q_{a} = 2 a \log |z|$
\begin{align}
T=- \frac \nu2  (\partial_z Q_{a})^2 - \mu_H \partial_z^2
Q_{a}=\frac{h_a}{z^2}, \la{Tflux}
\end{align}
where \(h_a\) is the conformal dimension \eq{8}. 

Finally, when the flux tube sits on top of a conical singularity, the stress tensor is the sum of \eq{Tflux} and {\eq{256}}. Near the singularity $T \sim ( \gamma \Delta_{\alpha} + h_{a}) / z^{2}$. { This}  implies the relation \eq{172}.  
 \smallskip

\noindent{\bf Exchange statistics} Now  consider  adiabatically exchanging  two singularities. The state will acquire a phase.
 Since the state is a holomorphic function of singularity position,  its holonomy is encoded in the normalization factor. The phase is then
  \(\Phi_{12}=\frac \ii 2\oint  d\log
\mathcal{Z}\), where the integral  in  positions of singularities goes along the adiabatic path.   The  adiabatic connection \(d\log \mathcal{Z}\) treated as a differential  of  the position, say, the first singularity has a pole when two singularities coincide, so the phase is the residue  of the pole { \(\Phi_{12}=-\pi\,{\rm res} [d\log \mathcal{Z}] \). 

 For conical singularities, the residue arises entirely from the gravitational
anomaly. We notice that  the calculation of the normalization factor for multiple singularities is closely related  to  the determinant of the Laplacian ${\rm Det (-\Delta)}$ on singular surfaces. This relation was discussed in 
\cite{CLWBig,lcw,KW2015}.  A reason for this is that the stress tensor for  $\log \mathcal{Z}$
 and $\frac{1}{2} c_{H}\log{\rm Det}(-\Delta)$, share the same singularities. 
 
 The result is summarized by the formula
\begin{align}
\log \mathcal{Z}|_{p_1\to p_2} =  \frac{c_{H}}{12}  \alpha_{1}\alpha_{2} \left( \frac{1}{\gamma_{1}} + \frac{1}{\gamma_{2}}\right) \log |p_1- p_2|.\la{38}
\end{align}
where  $p_{1} $ and $p_2$ are positions of two merging singularities.

Then the  adiabatic connection is 
\begin{align}
d \log \mathcal{Z} = \frac{c_{H}}{24} \alpha_{1}  \alpha_{2} \left( \frac{1}{\gamma_{1}} + \frac{1}{\gamma_{2}}\right) \frac{d p_1-dp_2}{p_1 - p_{2}}.
\end{align}
It  prompts the formula  \eq{10} for the exchange statistics.

We illustrate the calculation of  the generating functional  on the example of {{a genus-0 polyhedral surface}, such as a cube, tetrahedral, etc., whose vertices are separated by distances well exceeding the magnetic length.
 In this calculation we focus on the geometric singularities  setting the  fluxes  $a=0$.  The metric describing a polyhedron is piece-wise flat with multiple conical singularities.  It is obtained from the Schwarz-Christoffel  map $\xi(z) $ unfolding the polyhedron } \begin{align}
e^{\phi} = |{\xi(z)}'|^{2} = \prod_{i} |z - p_i |^{-2\alpha_{i}}. 
\end{align}
The metric describes a flat surface with conical singularities of the order $\alpha_i$, conditioned by the Gauss-Bonnet theorem  $-\sum_{i} \alpha_{i} + 2 = 0$, and located at points $p_i$. The magnetic  potential \(-\hbar\Delta Q=2eB\) corresponding to a uniform magnetic field will be $Q = - |\xi(z)|^{2} / 2l^{2}$. 

The Schwarzian of this metric is
\begin{align*}
\mathcal{S}[\phi] = \sum_{i} \frac{ - \frac{1}{2}\alpha_{i}^{2} + \alpha_{i}}{(z - p_i)^{2}} + \frac{\gamma_{i}}{z - p_i} , \quad \gamma_{i} \equiv - \sum_{j\ne i} \frac{\alpha_{i}\alpha_{j}}{p_i - p_j}.
\end{align*}
Using the explicit form of the metric, we find the following asymptotic behavior of the magnetic potential and the metric at  singularities 
\begin{align}
\partial_{p_i} \phi |_{z\to p_i}& = - \partial_{z}\phi  + \frac{\gamma_{i}}{\alpha_{i}} + ... ,\nn\\
\partial_{p_i} \phi |_{z= p_j}& =-\frac{\alpha_i}{p_i-p_j},\nn\\
\partial_{p_i} Q |_{z\to p_i}&=  (- \partial_{z}  + \frac{\gamma_{i}}{\alpha_{i}} )Q|_{z\to p_i} + ...,\la{41}\\
\partial_{p_i} Q|_{z= p_j}& =-\frac{\alpha_i}{p_i-p_j}Q|_{z= p_j}\nn
\end{align}
%
%
%
 Next, we use the formula    for the   angular momentum of a  that for a single  cone \eq{171}, which we write in the form
\begin{align}
\int_{D_{\epsilon}(p_{i})} \left(-Q + (j-1 ) \right)\left(  \rho  -\rho_{\infty}\right)dV & = \Delta_{\alpha_{i}}.
\la{42}\end{align}
To make sense of this formula for multiple cones, we take the domain of integration $D_{\epsilon}(p_{i})$ to be a small disk of radius $\epsilon$ centered on $p_{i}$ which is much larger than the magnetic length but much smaller than the Euclidean distance to the closest neighboring cone point.
In this case, $\rho_{\infty} = \frac{\nu}{2\pi l^{2}}$ is the density far from any cone points. 

For multiple cones, there is also another important non-vanishing sum rule that comes from the examining the simple poles of the Ward identity, which come entirely from the Schwarzian of the metric. This sum rule reads 
\begin{align}
	\int_{D_{\epsilon}(p_i)} \partial_{z} \left(  Q  + (1 -j) \phi \right) \left(  \rho  - \rho_{\infty}\right) dV = - \frac{c_{H}}{12} \gamma_{i}.\la{43}
\end{align}
 The proof for these sum rules comes directly from the Ward identity \eq{27}  specified for a uniform  magnetic field
\begin{align}
\frac{1}{\hbar} \int \frac{i \langle P_{z'} \rangle }{z - z'} dV_{z'} = \langle T \rangle  = \frac{c_{H}}{12} \mathcal{S}[\phi] + O(l^{2}).
\end{align}
Since the RHS is proportional to the Schwarzian, it will have second order pole and simple poles at the cone points. This implies that the integral on the LHS has a Laurent expansion around each cone point. Comparing the residues of the poles, we find \eq{42} comes from the second order pole of the Schwarzian, whereas \eq{43} follows from the simple pole.

Now we are equipped to derive the variational formula for the generating functional.

 Derivatives with respect to the cone points will act only the single-particle factors of the wave function \eq{12}, and thus lead to 
\begin{align}
\partial_{p_i} \log Z 
& = \int \partial_{p_i} \left( Q + (1 - j) \phi \right)  \rho ) dV.
\end{align}
In the integrand we write $\rho=(\rho-\rho_\infty)+\rho_\infty$.  The contribution of  $\rho_\infty$ is of the  order $l^{-2}$. We ignore it and focus on the contribution of $\rho-\rho_\infty$  which has  a finite support  at the conical singularity points $p_i$. Therefore, we convert the integral over the entire surface to a sum of integrals over  small disks $D_{\epsilon}(p_j)$ centered on singularities 
\begin{align*}
\partial_{p_i} \log Z = \sum_{j}
\int_{D_{\epsilon}(p_{j})}	\partial_{p_i} \left( Q + (1 - j) \phi \right) \left(  \rho  - \rho_{\infty}\right) dV.
\end{align*}
{  With this, we utilize the asymptotic expressions for $Q$ and $\phi$ in (41), combined with the formula \eq{42} for the moment of inertia and the  sum rule \eq{43}. }We  obtain 
\begin{align*}
\partial_{p_i} \log Z&= \frac{c_{H}}{12} \gamma_{i} - \frac{\gamma_{i}}{\alpha_{i}} \Delta_{\alpha_{i}}  + \sum_{j \ne i} \left(  \frac{\alpha_{i}}{p_i - p_j}\right) \Delta_{\alpha_{j}}\\
& =  \frac{c_{H}}{24} \sum_{j \ne i} \frac{\alpha_{i}\alpha_{j}}{p_i - p_j}\left[ \frac{ 1}{\gamma_{i}}  + \frac{ 1}{\gamma_{j} }\right].\nn
\end{align*}
This formula represents the adiabatic connection with respect to adiabatic displacement of  conical singularity announced  in the main text in Eq (30). 
 \smallskip

\noindent{\bf Integer QH-state on a cone} The formulae for the charge of the singularity
\eq{1311} and the moment of inertia \eq{11}
are readily  checked against the direct calculations for the integer case \(\nu=1\). See \cite{Furtado,*Furtado2,*Poux} for a study of Landau levels on a cone.
 In  the case where a flux  \( (h/ e)a \)  threads the cone  the Landau level is spanned by one-particle states
  $k=0,\dots, N\!-\!1 $
  \begin{equation}\la{321}
 \psi_k\!=\!\frac{e^{-{|\xi|}^2/{4l^2}}}{l\sqrt{2\pi\gamma\Gamma( \frac k\gamma+\frac q\gamma+1)}}\left(\frac{|\xi|}{
\sqrt{2} l}\right)^{ \frac q \gamma }\cdot \left(\!\!\frac\xi{
\sqrt{2} l}\!\right)^{ \frac k\gamma},
 \end{equation} 
 where  $q=a+\alpha j$,
%
and the density is the sum of
   densities  of each one-particle state \begin{align}\rho=\sum_{k=0}^{N-1}|\psi_k|^2.\la{90}
   \end{align}   
We observe that  in the integer  case  the magnetic singularity and spin come together in a combination  $q=a+\alpha j$. 
 \begin{figure}[htbp]
   \includegraphics[scale=0.255]{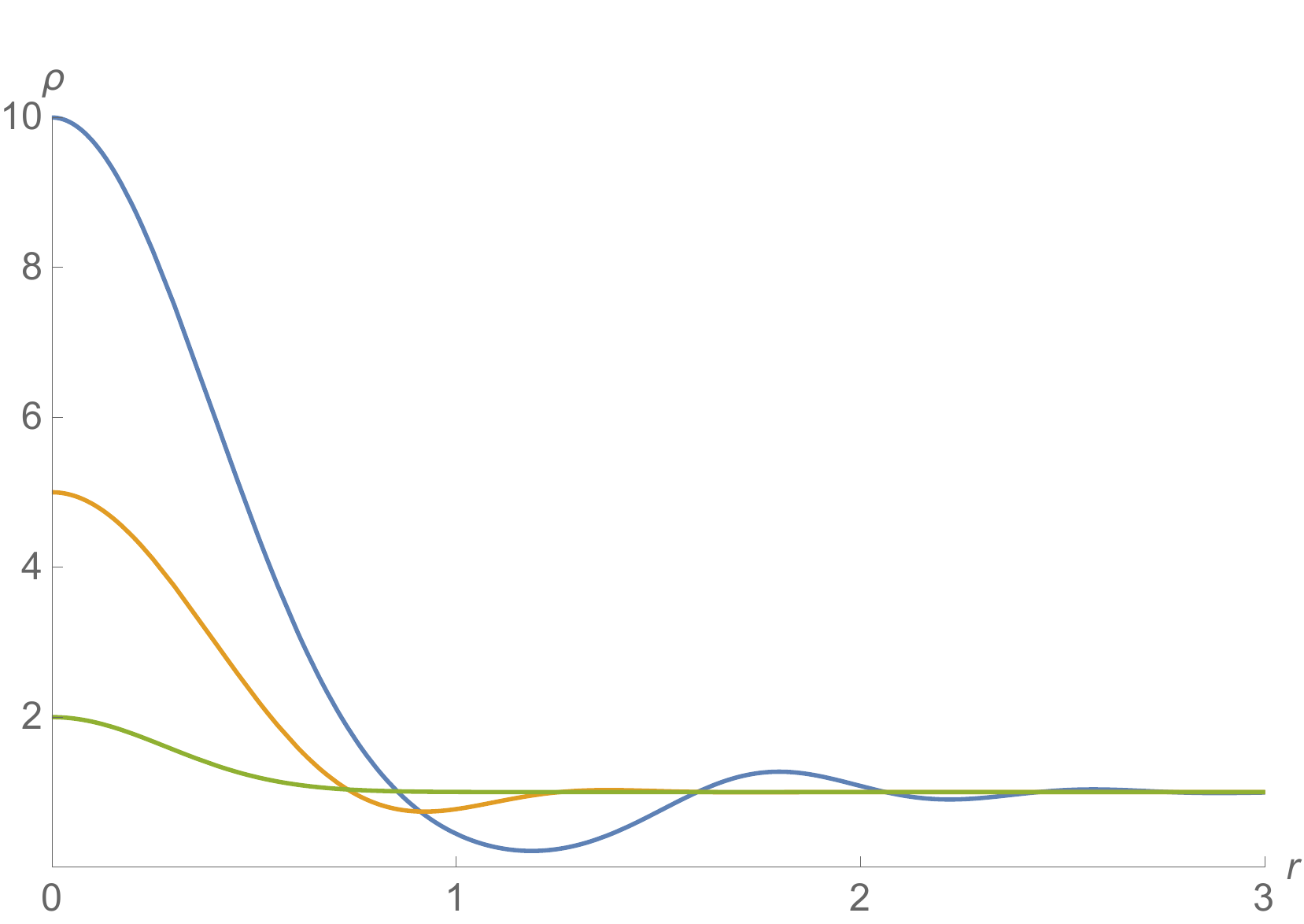}
 %
\includegraphics[scale=0.255]{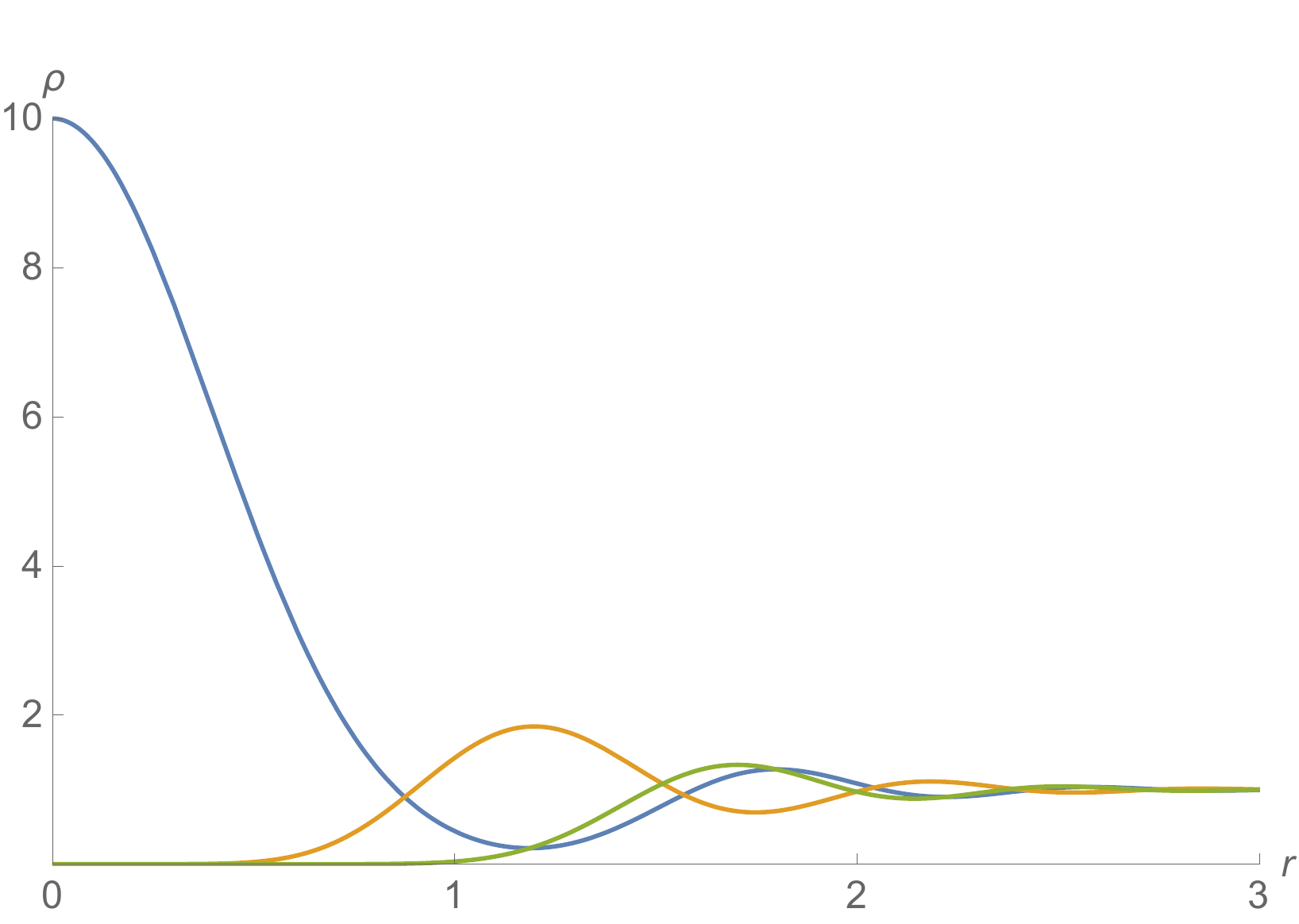}
   \caption{  Left panel: Density on a cone of angle $\gamma=1/10,1/5,1/2$ (blue, yellow, green) with spin  $j=0$. Right panel: Density  on a cone of angle $\gamma = 1/10$ and spin $j =0,1/2,1$ (blue, yellow, green). }
      \label{Fig2}
   \end{figure}

At $N\to \infty$  the density  is expressed in terms of Mittag-Leffler function 
 \be\nn
 \rho=\rho_{\infty}\gamma^{-1}e^{-x}x^{\frac q \gamma}E_{\frac 1\gamma, \frac q\gamma+1}\left(x^{\frac 1\gamma}\right),
 \ee
 where we denoted $x=|\xi|^2/2l^2$, and $\rho_{\infty} = 1/2\pi l^{2}$. We recall the definition of the Mittag-Leffler function $$E_{a,b}(z)=\sum_{k\geq 0}\frac{z^k}{\Gamma(ak+b)}.$$
 
 We find the moments from the arguments used to obtain the  conformal Ward identity. Under re-scaling
the magnetic length \(l^2\to\lambda^{-1} l^2\) the state \eq{321} scales $$\psi_k\to
\lambda^{\frac{1}{2}+\frac{k+q}{2\gamma}} e^{(1-\lambda){|\xi|}^2/{4l^2}}
\psi_k.$$
but remains normalized. The normalization condition for the new state yields the identity $$ \int e^{(1-\lambda)\frac{|\xi|^2}{2l^2}}|\psi_k|^2dV=\lambda^{-1 - (q + k)/\gamma}.$$
 Then, summing over all modes   and taking $N \to \infty$ at  $|\lambda|>1$ we obtain the exact Laplace transform of the density  
 \begin{align*}\int e^{(1-\lambda)\frac{|\xi|^2}{2l^2}}
(\rho-\rho_\infty) dV &=\frac{\lambda^{-1 - \frac{q}{\gamma} }}{1-\lambda^{-1/\gamma}}-\frac \gamma{\lambda-1}\\
& = \frac{\lambda^{\frac{\alpha - 2}{2\gamma} + \frac{m_{0}}{\gamma} }}{1-\lambda^{-1/\gamma}}-\frac \gamma{\lambda-1}
\end{align*}
where $m_{0} = - a + (1 - 2j) \alpha/2$ is the charge of the cone. This formula  can be seen as  a generating function of moments \eq{81}. 
Expanding  around \(\lambda=1\) yields 
 the charge \(m_0\)  \eq{1311} and   the   moment of inertia  \(m_2\)   \eq{11}.

In the orbifold setting the density is a finite  sum \cite{note1}
where $\gamma(s, x) = \int_{0}^{x} t^{s - 1}e^{ - t} dt$ is the lower incomplete gamma function. }At $\gamma=1/n$ and $q = 0$, the density is Riccati's {\it generalized hyperbolic function} - the sum over $n$-roots of unity $\omega^k_n=e^{ \ii\frac{2k\pi}{n}}$
    \begin{align*}
   \rho =\rho_\infty e^{ - x} \sum_{k= 0}^{n-1} \exp\left( \omega_n^{k}x \right)
   \end{align*}
 In Fig.\ref{Fig2}, we illustrate  how   the cone angle  and spin  affect the density, respectively. In both figures, the density far away from the singularity   is normalized to $1$ and the distance to the  conical point is measured in units of magnetic length.   
On the left panel  we set $j=0$ and plot $\rho$ for $\gamma = 1/2,1/5,1/10$. The values of density near the origin reflect the charge $m_0$ accumulated at the tip of the cone and feature oscillations on the order of magnetic length  away from the apex. We comment that magnetic singularity does not feature oscillations at $\nu=1$.
On the right panel we set $\gamma= 1/10$ and  show the effect of spin at $j=0,1/2, 2$.  Unless spin is zero the charge $m_0$ is negative.  The cone repels electrons. Spin also suppresses oscillations, but for  small values of spin oscillations  persist. For larger values of spin (not plotted) oscillations are  suppressed  entirely. 
\smallskip

Apart from the flat cone, there exist exact results for the ``American football" geometry, Fig. \ref{Fig3}, a unique surface with positive constant curvature and two conical singularities \cite{Troyanov1989}. The singularities have the same order and antipodal. The football  metric reads
\begin{align}
	ds^{2} =\frac{(1 -\alpha)^{2}  |z|^{-2\alpha}}{\left(1 + |z|^{2\gamma}/4 r^{2} \right)^{2}} |dz|^{2}, \quad \gamma=1-\alpha.
\end{align}
\begin{figure}[htbp]
\vspace{-4cm}
\begin{center}
\includegraphics[width=9.5cm]{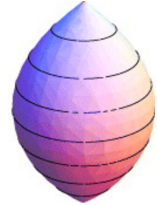}
\vspace{-4cm}
\caption{``American football": a surface of constant positive curvature and two antipodal conical singularities}
\label{Fig3}
\end{center}
\end{figure}
The volume and the curvature of this surface are\\ $V =  4 \pi r^{2} \gamma$ and $R = 2/r^{2}$, respectively. Identical conical singularities of order $\alpha$ are located at $z = 0$ and $z = \infty$.  The number of states is 
$
N = N_\Phi + (1 - 2j) - 2a
$,
where $N_\Phi $,  is the total magnetic flux, $a$ is the flux threaded through the singularities, and\begin{align*}
	\rho_{\infty} =\frac e hB + \frac{(1 - 2j)}{8\pi} \frac{2}{r^{2}} ,
\end{align*}
is the bulk density away from the conical singularities.

In coordinates $\xi = z^\gamma/(2r\gamma)$, the normalized eigenstates are 
\begin{align*}
\psi_{k} &= \frac{1}{\sqrt{V \mathcal{N}_{k}} }  \frac{\xi^{\frac{1}{\gamma}( k + q)}}{( 1 + |\xi|^{2})^{\frac{1}{2\gamma}{N}_\Phi- j} },	\quad q=a + \alpha j,
\end{align*}
where the normalization factor is 
\begin{align*}
\mathcal{N}_{k} & = B\left(\frac{1}{\gamma} N_{\Phi} - 2j + 1 - \frac 1{\gamma}{(k+q)} , \frac 1{\gamma}{(k + q)}+ 1\right)
\end{align*}
 and $B(x, y) = \Gamma(x) \Gamma(y)/\Gamma(x+y)$ is the beta-function.
 The formula \eq{90} gives the density
  \begin{align}
  \rho=&\rho_{\infty} \gamma^{-1} ( 1 + |\xi|^{2})^{-2 (N_{\Phi}/\gamma  - 2j)} \times \nn\\
 & \sum_{k=0}^{N-1} 
  \frac{\Gamma( N_{\Phi}/\gamma - 2j + 1)  |\xi|^{\frac{2}{\gamma}( k + q)}} {\Gamma( N_{\Phi}/\gamma - 2j + 1 - (k+q)/\gamma )\Gamma( (k + q)/\gamma+ 1)} .
 \end{align}
 As before, the density simplifies when $q = 0$ and $\gamma^{-1} = n$ is integer. Writing as before the $n^{th}$ root of unity $\omega_{n} = e^{ \ii 2\pi /n}$, we find
\begin{align}
\rho = \rho_{\infty} \left( 1 + |\xi|^{2}\right)^{-n N_{\Phi}} \sum_{k= 0}^{n-1} \left( 1 + \omega_{n}^{k}  |\xi|^{2} \right)^{ n N_{\Phi}}.
\end{align}
The charge moment follows from simple integration
\begin{align*}
	m_{0} &= \int \left(\langle \rho  \rangle - \rho_{\infty}\right)dV = N - N_\Phi + (1 - 2j)(1-\alpha)\\
	& = - (1- 2j) \alpha - 2a.
\end{align*}
This is twice the charge of a single flat cone given by \eq{1311}.

\bigskip


\noindent We thank A. Abanov, R. Biswas, S. Ganeshan, A. Gromov, S. Klevtsov, J. Ross, J. Simon,  D. T.
Son, A. Waldron, and S. Zelditch for illuminating discussions. The work  was supported in part  by the NSF under the  Grants
NSF DMR-1206648, NSF DMR-MRSEC 1420709, and NSF PHY11-25915.

\bibliography{conepaper}

\begin{thebibliography}{45}%
\makeatletter
\providecommand \@ifxundefined [1]{%
 \@ifx{#1\undefined}
}%
\providecommand \@ifnum [1]{%
 \ifnum #1\expandafter \@firstoftwo
 \else \expandafter \@secondoftwo
 \fi
}%
\providecommand \@ifx [1]{%
 \ifx #1\expandafter \@firstoftwo
 \else \expandafter \@secondoftwo
 \fi
}%
\providecommand \natexlab [1]{#1}%
\providecommand \enquote  [1]{``#1''}%
\providecommand \bibnamefont  [1]{#1}%
\providecommand \bibfnamefont [1]{#1}%
\providecommand \citenamefont [1]{#1}%
\providecommand \href@noop [0]{\@secondoftwo}%
\providecommand \href [0]{\begingroup \@sanitize@url \@href}%
\providecommand \@href[1]{\@@startlink{#1}\@@href}%
\providecommand \@@href[1]{\endgroup#1\@@endlink}%
\providecommand \@sanitize@url [0]{\catcode `\\12\catcode `\$12\catcode
  `\&12\catcode `\#12\catcode `\^12\catcode `\_12\catcode `\%12\relax}%
\providecommand \@@startlink[1]{}%
\providecommand \@@endlink[0]{}%
\providecommand \url  [0]{\begingroup\@sanitize@url \@url }%
\providecommand \@url [1]{\endgroup\@href {#1}{\urlprefix }}%
\providecommand \urlprefix  [0]{URL }%
\providecommand \Eprint [0]{\href }%
\providecommand \doibase [0]{http://dx.doi.org/}%
\providecommand \selectlanguage [0]{\@gobble}%
\providecommand \bibinfo  [0]{\@secondoftwo}%
\providecommand \bibfield  [0]{\@secondoftwo}%
\providecommand \translation [1]{[#1]}%
\providecommand \BibitemOpen [0]{}%
\providecommand \bibitemStop [0]{}%
\providecommand \bibitemNoStop [0]{.\EOS\space}%
\providecommand \EOS [0]{\spacefactor3000\relax}%
\providecommand \BibitemShut  [1]{\csname bibitem#1\endcsname}%
\let\auto@bib@innerbib\@empty
\bibitem [{\citenamefont {Avron}\ \emph {et~al.}(1995)\citenamefont {Avron},
  \citenamefont {Seiler},\ and\ \citenamefont {Zograf}}]{Seiler1995}%
  \BibitemOpen
  \bibfield  {author} {\bibinfo {author} {\bibfnamefont {J.~E.}\ \bibnamefont
  {Avron}}, \bibinfo {author} {\bibfnamefont {R.}~\bibnamefont {Seiler}}, \
  and\ \bibinfo {author} {\bibfnamefont {P.~G.}\ \bibnamefont {Zograf}},\
  }\href@noop {} {\bibfield  {journal} {\bibinfo  {journal} {Phys. Rev. Lett.}\
  }\textbf {\bibinfo {volume} {75}},\ \bibinfo {pages} {697} (\bibinfo {year}
  {1995})}\BibitemShut {NoStop}%
\bibitem [{\citenamefont {L{\'e}vay}(1995)}]{Levay}%
  \BibitemOpen
  \bibfield  {author} {\bibinfo {author} {\bibfnamefont {P.}~\bibnamefont
  {L{\'e}vay}},\ }\href {\doibase http://dx.doi.org/10.1063/1.531066}
  {\bibfield  {journal} {\bibinfo  {journal} {J. Math. Phys.}\ }\textbf
  {\bibinfo {volume} {36}},\ \bibinfo {pages} {2792} (\bibinfo {year}
  {1995})}\BibitemShut {NoStop}%
\bibitem [{\citenamefont {L\'evay}(1997)}]{Levay1997}%
  \BibitemOpen
  \bibfield  {author} {\bibinfo {author} {\bibfnamefont {P.}~\bibnamefont
  {L\'evay}},\ }\href {\doibase 10.1103/PhysRevE.56.6173} {\bibfield  {journal}
  {\bibinfo  {journal} {Phys. Rev. E}\ }\textbf {\bibinfo {volume} {56}},\
  \bibinfo {pages} {6173} (\bibinfo {year} {1997})}\BibitemShut {NoStop}%
\bibitem [{\citenamefont {Wen}\ and\ \citenamefont {Zee}(1992)}]{Wen1992a}%
  \BibitemOpen
  \bibfield  {author} {\bibinfo {author} {\bibfnamefont {X.~G.}\ \bibnamefont
  {Wen}}\ and\ \bibinfo {author} {\bibfnamefont {A.}~\bibnamefont {Zee}},\
  }\href {\doibase 10.1103/PhysRevLett.69.953} {\bibfield  {journal} {\bibinfo
  {journal} {Phys. Rev. Lett.}\ }\textbf {\bibinfo {volume} {69}},\ \bibinfo
  {pages} {953} (\bibinfo {year} {1992})}\BibitemShut {NoStop}%
\bibitem [{\citenamefont {Fr{\"o}hlich}\ and\ \citenamefont
  {Studer}(1992)}]{Frohlich1992}%
  \BibitemOpen
  \bibfield  {author} {\bibinfo {author} {\bibfnamefont {J.}~\bibnamefont
  {Fr{\"o}hlich}}\ and\ \bibinfo {author} {\bibfnamefont {U.~M.}\ \bibnamefont
  {Studer}},\ }\href@noop {} {\bibfield  {journal} {\bibinfo  {journal}
  {Commun. Math. Phys.}\ }\textbf {\bibinfo {volume} {148}},\ \bibinfo {pages}
  {553} (\bibinfo {year} {1992})}\BibitemShut {NoStop}%
\bibitem [{\citenamefont {Hoyos}\ and\ \citenamefont {Son}(2012)}]{HoyosSon}%
  \BibitemOpen
  \bibfield  {author} {\bibinfo {author} {\bibfnamefont {C.}~\bibnamefont
  {Hoyos}}\ and\ \bibinfo {author} {\bibfnamefont {D.~T.}\ \bibnamefont
  {Son}},\ }\href {\doibase 10.1103/PhysRevLett.108.066805} {\bibfield
  {journal} {\bibinfo  {journal} {Phys. Rev. Lett.}\ }\textbf {\bibinfo
  {volume} {108}},\ \bibinfo {pages} {066805} (\bibinfo {year} {2012})},\
  \bibinfo {note} {\href{http://arxiv.org/abs/1109.2651} {\tt
  arXiv:1109.2651}}\BibitemShut {NoStop}%
\bibitem [{\citenamefont {Read}(2009)}]{Read2009}%
  \BibitemOpen
  \bibfield  {author} {\bibinfo {author} {\bibfnamefont {N.}~\bibnamefont
  {Read}},\ }\href {\doibase 10.1103/PhysRevB.79.045308} {\bibfield  {journal}
  {\bibinfo  {journal} {Phys. Rev. B}\ }\textbf {\bibinfo {volume} {79}},\
  \bibinfo {pages} {045308} (\bibinfo {year} {2009})},\ \bibinfo {note}
  {\href{http://arxiv.org/abs/0805.2507}{\tt arXiv:0805.2507v3}}\BibitemShut
  {NoStop}%
\bibitem [{\citenamefont {Read}\ and\ \citenamefont {Rezayi}(2011)}]{Read2011}%
  \BibitemOpen
  \bibfield  {author} {\bibinfo {author} {\bibfnamefont {N.}~\bibnamefont
  {Read}}\ and\ \bibinfo {author} {\bibfnamefont {E.~H.}\ \bibnamefont
  {Rezayi}},\ }\href {\doibase 10.1103/PhysRevB.84.085316} {\bibfield
  {journal} {\bibinfo  {journal} {Phys. Rev. B}\ }\textbf {\bibinfo {volume}
  {84}},\ \bibinfo {pages} {085316} (\bibinfo {year} {2011})},\ \bibinfo {note}
  {\href{http://arxiv.org/abs/1008.0210} {\tt arXiv:1008.0210}}\BibitemShut
  {NoStop}%
\bibitem [{\citenamefont {Douglas}\ and\ \citenamefont
  {Klevtsov}(2009)}]{Douglas2010}%
  \BibitemOpen
  \bibfield  {author} {\bibinfo {author} {\bibfnamefont {M.~R.}\ \bibnamefont
  {Douglas}}\ and\ \bibinfo {author} {\bibfnamefont {S.}~\bibnamefont
  {Klevtsov}},\ }\href {\doibase 10.1007/s00220-009-0915-0} {\bibfield
  {journal} {\bibinfo  {journal} {Commun. Math. Phys.}\ }\textbf {\bibinfo
  {volume} {293}},\ \bibinfo {pages} {205} (\bibinfo {year} {2009})},\ \bibinfo
  {note} {\href{http://arxiv.org/abs/0808.2451}{\tt
  arXiv:0808.2451v1}}\BibitemShut {NoStop}%
\bibitem [{\citenamefont {Klevtsov}(2014)}]{Klevtsov2013}%
  \BibitemOpen
  \bibfield  {author} {\bibinfo {author} {\bibfnamefont {S.}~\bibnamefont
  {Klevtsov}},\ }\href {\doibase 10.1007/JHEP01(2014)133} {\bibfield  {journal}
  {\bibinfo  {journal} {J. High Energy Phys.}\ }\textbf {\bibinfo {volume}
  {2014}},\ \bibinfo {pages} {1} (\bibinfo {year} {2014})},\ \bibinfo {note}
  {\href{http://arxiv.org/abs/1309.7333}{\tt arXiv:1309.7333v2}}\BibitemShut
  {NoStop}%
\bibitem [{\citenamefont {Can}\ \emph {et~al.}(2014)\citenamefont {Can},
  \citenamefont {Laskin},\ and\ \citenamefont {Wiegmann}}]{CLW}%
  \BibitemOpen
  \bibfield  {author} {\bibinfo {author} {\bibfnamefont {T.}~\bibnamefont
  {Can}}, \bibinfo {author} {\bibfnamefont {M.}~\bibnamefont {Laskin}}, \ and\
  \bibinfo {author} {\bibfnamefont {P.}~\bibnamefont {Wiegmann}},\ }\href
  {\doibase 10.1103/PhysRevLett.113.046803} {\bibfield  {journal} {\bibinfo
  {journal} {Phys. Rev. Lett.}\ }\textbf {\bibinfo {volume} {113}},\ \bibinfo
  {pages} {046803} (\bibinfo {year} {2014})},\ \bibinfo {note}
  {\href{http://arxiv.org/abs/1402.1531}{\tt arXiv:1402.1531v2}}\BibitemShut
  {NoStop}%
\bibitem [{\citenamefont {Can}\ \emph {et~al.}(2015)\citenamefont {Can},
  \citenamefont {Laskin},\ and\ \citenamefont {Wiegmann}}]{CLWBig}%
  \BibitemOpen
  \bibfield  {author} {\bibinfo {author} {\bibfnamefont {T.}~\bibnamefont
  {Can}}, \bibinfo {author} {\bibfnamefont {M.}~\bibnamefont {Laskin}}, \ and\
  \bibinfo {author} {\bibfnamefont {P.~B.}\ \bibnamefont {Wiegmann}},\ }\href
  {\doibase http://dx.doi.org/10.1016/j.aop.2015.02.013} {\bibfield  {journal}
  {\bibinfo  {journal} {Ann. Phys.}\ }\textbf {\bibinfo {volume} {362}},\
  \bibinfo {pages} {752 } (\bibinfo {year} {2015})},\ \bibinfo {note}
  {\href{http://arxiv.org/abs/1411.3105}{\tt arXiv:1411.3105v3}}\BibitemShut
  {NoStop}%
\bibitem [{\citenamefont {Abanov}\ and\ \citenamefont
  {Gromov}(2014)}]{Abanov2014}%
  \BibitemOpen
  \bibfield  {author} {\bibinfo {author} {\bibfnamefont {A.~G.}\ \bibnamefont
  {Abanov}}\ and\ \bibinfo {author} {\bibfnamefont {A.}~\bibnamefont
  {Gromov}},\ }\href {\doibase 10.1103/PhysRevB.90.014435} {\bibfield
  {journal} {\bibinfo  {journal} {Phys. Rev. B}\ }\textbf {\bibinfo {volume}
  {90}},\ \bibinfo {pages} {014435} (\bibinfo {year} {2014})},\ \bibinfo {note}
  {\href{http://arxiv.org/abs/1401.3703}{\tt arXiv:1401.3703v1}}\BibitemShut
  {NoStop}%
\bibitem [{\citenamefont {Gromov}\ and\ \citenamefont
  {Abanov}(2014)}]{GromovAbanov}%
  \BibitemOpen
  \bibfield  {author} {\bibinfo {author} {\bibfnamefont {A.}~\bibnamefont
  {Gromov}}\ and\ \bibinfo {author} {\bibfnamefont {A.~G.}\ \bibnamefont
  {Abanov}},\ }\href {\doibase 10.1103/PhysRevLett.113.266802} {\bibfield
  {journal} {\bibinfo  {journal} {Phys. Rev. Lett.}\ }\textbf {\bibinfo
  {volume} {113}},\ \bibinfo {pages} {266802} (\bibinfo {year} {2014})},\
  \bibinfo {note} {\href{http://arxiv.org/abs/1403.5809}{\tt
  arXiv:1403.5809v1}}\BibitemShut {NoStop}%
\bibitem [{\citenamefont {Gromov}\ \emph {et~al.}(2015)\citenamefont {Gromov},
  \citenamefont {Cho}, \citenamefont {You}, \citenamefont {Abanov},\ and\
  \citenamefont {Fradkin}}]{framinganomaly}%
  \BibitemOpen
  \bibfield  {author} {\bibinfo {author} {\bibfnamefont {A.}~\bibnamefont
  {Gromov}}, \bibinfo {author} {\bibfnamefont {G.~Y.}\ \bibnamefont {Cho}},
  \bibinfo {author} {\bibfnamefont {Y.}~\bibnamefont {You}}, \bibinfo {author}
  {\bibfnamefont {A.~G.}\ \bibnamefont {Abanov}}, \ and\ \bibinfo {author}
  {\bibfnamefont {E.}~\bibnamefont {Fradkin}},\ }\href {\doibase
  10.1103/PhysRevLett.114.016805} {\bibfield  {journal} {\bibinfo  {journal}
  {Phys. Rev. Lett.}\ }\textbf {\bibinfo {volume} {114}},\ \bibinfo {pages}
  {016805} (\bibinfo {year} {2015})},\ \bibinfo {note}
  {\href{http://arxiv.org/abs/1410.6812}{\tt arXiv:1410.6812v3}}\BibitemShut
  {NoStop}%
\bibitem [{\citenamefont {Laskin}\ \emph {et~al.}(2015)\citenamefont {Laskin},
  \citenamefont {Can},\ and\ \citenamefont {Wiegmann}}]{lcw}%
  \BibitemOpen
  \bibfield  {author} {\bibinfo {author} {\bibfnamefont {M.}~\bibnamefont
  {Laskin}}, \bibinfo {author} {\bibfnamefont {T.}~\bibnamefont {Can}}, \ and\
  \bibinfo {author} {\bibfnamefont {P.}~\bibnamefont {Wiegmann}},\ }\href
  {\doibase 10.1103/PhysRevB.92.235141} {\bibfield  {journal} {\bibinfo
  {journal} {Phys. Rev. B}\ }\textbf {\bibinfo {volume} {92}},\ \bibinfo
  {pages} {235141} (\bibinfo {year} {2015})},\ \bibinfo {note}
  {\href{http://arxiv.org/abs/1412.8716}{\tt arXiv:1412.8716v2}}\BibitemShut
  {NoStop}%
\bibitem [{\citenamefont {Ferrari}\ and\ \citenamefont
  {Klevtsov}(2014)}]{Klevtsov2014}%
  \BibitemOpen
  \bibfield  {author} {\bibinfo {author} {\bibfnamefont {F.}~\bibnamefont
  {Ferrari}}\ and\ \bibinfo {author} {\bibfnamefont {S.}~\bibnamefont
  {Klevtsov}},\ }\href {\doibase 10.1007/JHEP12(2014)086} {\bibfield  {journal}
  {\bibinfo  {journal} {J. High Energy Phys.}\ }\textbf {\bibinfo {volume}
  {2014}},\ \bibinfo {pages} {1} (\bibinfo {year} {2014})},\ \bibinfo {note}
  {\href{http://arxiv.org/abs/1410.6802}{\tt arXiv:1410.6802v3}}\BibitemShut
  {NoStop}%
\bibitem [{\citenamefont {Bradlyn}\ and\ \citenamefont
  {Read}(2015)}]{Read2015}%
  \BibitemOpen
  \bibfield  {author} {\bibinfo {author} {\bibfnamefont {B.}~\bibnamefont
  {Bradlyn}}\ and\ \bibinfo {author} {\bibfnamefont {N.}~\bibnamefont {Read}},\
  }\href {\doibase 10.1103/PhysRevB.91.165306} {\bibfield  {journal} {\bibinfo
  {journal} {Phys. Rev. B}\ }\textbf {\bibinfo {volume} {91}},\ \bibinfo
  {pages} {165306} (\bibinfo {year} {2015})},\ \bibinfo {note}
  {\href{http://arxiv.org/abs/1502.04126}{\tt arXiv:1502.04126v2}}\BibitemShut
  {NoStop}%
\bibitem [{\citenamefont {Klevtsov}\ and\ \citenamefont
  {Wiegmann}(2015)}]{KW2015}%
  \BibitemOpen
  \bibfield  {author} {\bibinfo {author} {\bibfnamefont {S.}~\bibnamefont
  {Klevtsov}}\ and\ \bibinfo {author} {\bibfnamefont {P.}~\bibnamefont
  {Wiegmann}},\ }\href {\doibase 10.1103/PhysRevLett.115.086801} {\bibfield
  {journal} {\bibinfo  {journal} {Phys. Rev. Lett.}\ }\textbf {\bibinfo
  {volume} {115}},\ \bibinfo {pages} {086801} (\bibinfo {year} {2015})},\
  \bibinfo {note} {\href{http://arxiv.org/abs/1504.07198}{\tt
  arXiv:1504.07198v2}}\BibitemShut {NoStop}%
\bibitem [{\citenamefont {{Vafa}}(2015)}]{Vafa}%
  \BibitemOpen
  \bibfield  {author} {\bibinfo {author} {\bibfnamefont {C.}~\bibnamefont
  {{Vafa}}},\ }\href@noop {} {\  (\bibinfo {year} {2015})},\ \bibinfo {note}
  {\href{http://arxiv.org/abs/1511.03372}{\tt arXiv:1511.03372v3}},\ \Eprint
  {http://arxiv.org/abs/1511.03372} {arXiv:1511.03372 [cond-mat.mes-{H}all]}
  \BibitemShut {NoStop}%
\bibitem [{\citenamefont {Katanaev}\ and\ \citenamefont
  {Volovich}(1992)}]{Katanaev1992}%
  \BibitemOpen
  \bibfield  {author} {\bibinfo {author} {\bibfnamefont {M.}~\bibnamefont
  {Katanaev}}\ and\ \bibinfo {author} {\bibfnamefont {I.}~\bibnamefont
  {Volovich}},\ }\href {\doibase
  http://dx.doi.org/10.1016/0003-4916(52)90040-7} {\bibfield  {journal}
  {\bibinfo  {journal} {Ann. Phys.}\ }\textbf {\bibinfo {volume} {216}},\
  \bibinfo {pages} {1 } (\bibinfo {year} {1992})}\BibitemShut {NoStop}%
\bibitem [{\citenamefont {Lammert}\ and\ \citenamefont
  {Crespi}(2004)}]{graphene1}%
  \BibitemOpen
  \bibfield  {author} {\bibinfo {author} {\bibfnamefont {P.~E.}\ \bibnamefont
  {Lammert}}\ and\ \bibinfo {author} {\bibfnamefont {V.~H.}\ \bibnamefont
  {Crespi}},\ }\href {\doibase 10.1103/PhysRevB.69.035406} {\bibfield
  {journal} {\bibinfo  {journal} {Phys. Rev. B}\ }\textbf {\bibinfo {volume}
  {69}},\ \bibinfo {pages} {035406} (\bibinfo {year} {2004})}\BibitemShut
  {NoStop}%
\bibitem [{\citenamefont {Vozmediano}\ \emph {et~al.}(2010)\citenamefont
  {Vozmediano}, \citenamefont {Katsnelson},\ and\ \citenamefont
  {Guinea}}]{graphene2}%
  \BibitemOpen
  \bibfield  {author} {\bibinfo {author} {\bibfnamefont {M.}~\bibnamefont
  {Vozmediano}}, \bibinfo {author} {\bibfnamefont {M.}~\bibnamefont
  {Katsnelson}}, \ and\ \bibinfo {author} {\bibfnamefont {F.}~\bibnamefont
  {Guinea}},\ }\href {\doibase http://dx.doi.org/10.1016/j.physrep.2010.07.003}
  {\bibfield  {journal} {\bibinfo  {journal} {Phys. Rep.}\ }\textbf {\bibinfo
  {volume} {496}},\ \bibinfo {pages} {109 } (\bibinfo {year}
  {2010})}\BibitemShut {NoStop}%
\bibitem [{\citenamefont {{Schine}}\ \emph {et~al.}(2015)\citenamefont
  {{Schine}}, \citenamefont {{Ryou}}, \citenamefont {{Gromov}}, \citenamefont
  {{Sommer}},\ and\ \citenamefont {{Simon}}}]{Simon2015}%
  \BibitemOpen
  \bibfield  {author} {\bibinfo {author} {\bibfnamefont {N.}~\bibnamefont
  {{Schine}}}, \bibinfo {author} {\bibfnamefont {A.}~\bibnamefont {{Ryou}}},
  \bibinfo {author} {\bibfnamefont {A.}~\bibnamefont {{Gromov}}}, \bibinfo
  {author} {\bibfnamefont {A.}~\bibnamefont {{Sommer}}}, \ and\ \bibinfo
  {author} {\bibfnamefont {J.}~\bibnamefont {{Simon}}},\ }\href@noop {} {\
  (\bibinfo {year} {2015})},\ \bibinfo {note}
  {\href{http://arxiv.org/abs/1511.07381}{\tt arXiv:1511.07381v1}},\ \Eprint
  {http://arxiv.org/abs/1511.07381} {arXiv:1511.07381 [cond-mat.quant-gas]}
  \BibitemShut {NoStop}%
\bibitem [{\citenamefont {Troyanov}(1989)}]{Troyanov1989}%
  \BibitemOpen
  \bibfield  {author} {\bibinfo {author} {\bibfnamefont {M.}~\bibnamefont
  {Troyanov}},\ }\href {\doibase 10.1007/BFb0086431} {\bibfield  {journal}
  {\bibinfo  {journal} {Lecture Notes in Math}\ }\textbf {\bibinfo {volume}
  {1410}},\ \bibinfo {pages} {296} (\bibinfo {year} {1989})}\BibitemShut
  {NoStop}%
\bibitem [{\citenamefont {Troyanov}(2007)}]{Troyanov2007}%
  \BibitemOpen
  \bibfield  {author} {\bibinfo {author} {\bibfnamefont {M.}~\bibnamefont
  {Troyanov}},\ }\href@noop {} {\bibfield  {journal} {\bibinfo  {journal} {IRMA
  Lectures in Math. and Theor. Phys.}\ }\textbf {\bibinfo {volume} {11}},\
  \bibinfo {pages} {507} (\bibinfo {year} {2007})},\ \bibinfo {note}
  {\href{http://arxiv.org/abs/math/0702666}{\tt arXiv:1504.07198}}\BibitemShut
  {NoStop}%
\bibitem [{\citenamefont {{Thurston}}(1998)}]{Thurston1998}%
  \BibitemOpen
  \bibfield  {author} {\bibinfo {author} {\bibfnamefont {W.~P.}\ \bibnamefont
  {{Thurston}}},\ }\href@noop {} {\bibfield  {journal} {\bibinfo  {journal}
  {Geom. Topol. Monogr.}\ }\textbf {\bibinfo {volume} {1}},\ \bibinfo {pages}
  {511} (\bibinfo {year} {1998})},\ \bibinfo {note}
  {\href{http://arxiv.org/abs/math/9801088}{\tt arXiv:math/9801088v2}},\
  \Eprint {http://arxiv.org/abs/math/9801088} {math/9801088} \BibitemShut
  {NoStop}%
\bibitem [{\citenamefont {Takhtajan}\ and\ \citenamefont
  {Zograf}(2002)}]{TZ2002}%
  \BibitemOpen
  \bibfield  {author} {\bibinfo {author} {\bibfnamefont {L.}~\bibnamefont
  {Takhtajan}}\ and\ \bibinfo {author} {\bibfnamefont {P.}~\bibnamefont
  {Zograf}},\ }\href@noop {} {\bibfield  {journal} {\bibinfo  {journal} {Trans.
  Amer. Math. Soc.}\ }\textbf {\bibinfo {volume} {355}},\ \bibinfo {pages}
  {1857} (\bibinfo {year} {2002})},\ \bibinfo {note}
  {\href{http://arxiv.org/abs/math/0112170}{\tt
  arXiv:math/0112170}}\BibitemShut {NoStop}%
\bibitem [{\citenamefont {Knizhnik}(1987)}]{Knizhnik1987}%
  \BibitemOpen
  \bibfield  {author} {\bibinfo {author} {\bibfnamefont {V.~G.}\ \bibnamefont
  {Knizhnik}},\ }\href {http://projecteuclid.org/euclid.cmp/1104160053}
  {\bibfield  {journal} {\bibinfo  {journal} {Commun. Math. Phys.}\ }\textbf
  {\bibinfo {volume} {112}},\ \bibinfo {pages} {567} (\bibinfo {year}
  {1987})}\BibitemShut {NoStop}%
\bibitem [{\citenamefont {Bershadsky}\ and\ \citenamefont
  {Radul}(1987)}]{Bershadsky1988}%
  \BibitemOpen
  \bibfield  {author} {\bibinfo {author} {\bibfnamefont {M.}~\bibnamefont
  {Bershadsky}}\ and\ \bibinfo {author} {\bibfnamefont {A.}~\bibnamefont
  {Radul}},\ }\href {\doibase 10.1142/S0217751X87000053} {\bibfield  {journal}
  {\bibinfo  {journal} {Int. J. Mod. Phys.}\ }\textbf {\bibinfo {volume}
  {A2}},\ \bibinfo {pages} {165} (\bibinfo {year} {1987})}\BibitemShut
  {NoStop}%
\bibitem [{\citenamefont {{Kvorning}}(2013)}]{Kvorning}%
  \BibitemOpen
  \bibfield  {author} {\bibinfo {author} {\bibfnamefont {T.}~\bibnamefont
  {{Kvorning}}},\ }\href {\doibase 10.1103/PhysRevB.87.195131} {\bibfield
  {journal} {\bibinfo  {journal} {\prb}\ }\textbf {\bibinfo {volume} {87}},\
  \bibinfo {eid} {195131} (\bibinfo {year} {2013})},\ \bibinfo {note}
  {\href{http://arxiv.org/abs/1302.3808}{\tt arXiv:1302.3808v2}},\ \Eprint
  {http://arxiv.org/abs/1302.3808} {arXiv:1302.3808 [cond-mat.str-el]}
  \BibitemShut {NoStop}%
\bibitem [{\citenamefont {Cardy}\ and\ \citenamefont
  {Peschel}(1988)}]{Cardy1988}%
  \BibitemOpen
  \bibfield  {author} {\bibinfo {author} {\bibfnamefont {J.~L.}\ \bibnamefont
  {Cardy}}\ and\ \bibinfo {author} {\bibfnamefont {I.}~\bibnamefont
  {Peschel}},\ }\href {\doibase http://dx.doi.org/10.1016/0550-3213(88)90604-9}
  {\bibfield  {journal} {\bibinfo  {journal} {Nucl. Phys. B}\ }\textbf
  {\bibinfo {volume} {300}},\ \bibinfo {pages} {377 } (\bibinfo {year}
  {1988})}\BibitemShut {NoStop}%
\bibitem [{\citenamefont {Kokotov}(2013)}]{Kokotov}%
  \BibitemOpen
  \bibfield  {author} {\bibinfo {author} {\bibfnamefont {A.}~\bibnamefont
  {Kokotov}},\ }\href {\doibase 10.1090/S0002-9939-2012-11531-X} {\bibfield
  {journal} {\bibinfo  {journal} {Proc. Math. Soc.}\ }\textbf {\bibinfo
  {volume} {141}},\ \bibinfo {pages} {725} (\bibinfo {year}
  {2013})}\BibitemShut {NoStop}%
\bibitem [{\citenamefont {Aurell}\ and\ \citenamefont
  {Salomonson}(1994)}]{Aurell2}%
  \BibitemOpen
  \bibfield  {author} {\bibinfo {author} {\bibfnamefont {E.}~\bibnamefont
  {Aurell}}\ and\ \bibinfo {author} {\bibfnamefont {P.}~\bibnamefont
  {Salomonson}},\ }\href@noop {} {\  (\bibinfo {year} {1994})},\ \bibinfo
  {note} {\href{http://arxiv.org/abs/hep-th/9405140}{\tt arXiv:9405140}},\
  \Eprint {http://arxiv.org/abs/hep-th/9405140} {arXiv:hep-th/9405140 [hep-th]}
  \BibitemShut {NoStop}%
\bibitem [{\citenamefont {Arovas}\ \emph {et~al.}(1984)\citenamefont {Arovas},
  \citenamefont {Schrieffer},\ and\ \citenamefont {Wilczek}}]{Arovas1984}%
  \BibitemOpen
  \bibfield  {author} {\bibinfo {author} {\bibfnamefont {D.}~\bibnamefont
  {Arovas}}, \bibinfo {author} {\bibfnamefont {J.~R.}\ \bibnamefont
  {Schrieffer}}, \ and\ \bibinfo {author} {\bibfnamefont {F.}~\bibnamefont
  {Wilczek}},\ }\href {\doibase 10.1103/PhysRevLett.53.722} {\bibfield
  {journal} {\bibinfo  {journal} {Phys. Rev. Lett.}\ }\textbf {\bibinfo
  {volume} {53}},\ \bibinfo {pages} {722} (\bibinfo {year} {1984})}\BibitemShut
  {NoStop}%
\bibitem [{\citenamefont {{Biswas}}\ and\ \citenamefont
  {{Son}}(2014)}]{Biswas2014}%
  \BibitemOpen
  \bibfield  {author} {\bibinfo {author} {\bibfnamefont {R.~R.}\ \bibnamefont
  {{Biswas}}}\ and\ \bibinfo {author} {\bibfnamefont {D.~T.}\ \bibnamefont
  {{Son}}},\ }\href@noop {} {\  (\bibinfo {year} {2014})},\ \bibinfo {note}
  {\href{http://arxiv.org/abs/1412.3809}{\tt arXiv:1412.3809v1}},\ \Eprint
  {http://arxiv.org/abs/1412.3809} {arXiv:1412.3809 [cond-mat.mes-{H}all]}
  \BibitemShut {NoStop}%
\bibitem [{\citenamefont {Wiegmann}(2013{\natexlab{a}})}]{Wiegmann2013}%
  \BibitemOpen
  \bibfield  {author} {\bibinfo {author} {\bibfnamefont {P.}~\bibnamefont
  {Wiegmann}},\ }\href {\doibase 10.1134/S1063776113110162} {\bibfield
  {journal} {\bibinfo  {journal} {J. Exp. Theor. Phys.}\ }\textbf {\bibinfo
  {volume} {117}},\ \bibinfo {pages} {538} (\bibinfo {year}
  {2013}{\natexlab{a}})},\ \bibinfo {note}
  {\href{http://arxiv.org/abs/1305.6893}{\tt arXiv:1305.6893v2}}\BibitemShut
  {NoStop}%
\bibitem [{\citenamefont {{Wiegmann}}(2012)}]{PW12}%
  \BibitemOpen
  \bibfield  {author} {\bibinfo {author} {\bibfnamefont {P.}~\bibnamefont
  {{Wiegmann}}},\ }\href@noop {} {\  (\bibinfo {year} {2012})},\ \bibinfo
  {note} {\href{http://arxiv.org/abs/1211.5132}{\tt arXiv:1211.5132v1}},\
  \Eprint {http://arxiv.org/abs/1211.5132} {arXiv:1211.5132 [cond-mat.str-el]}
  \BibitemShut {NoStop}%
\bibitem [{\citenamefont {Wiegmann}(2013{\natexlab{b}})}]{PW13}%
  \BibitemOpen
  \bibfield  {author} {\bibinfo {author} {\bibfnamefont {P.~B.}\ \bibnamefont
  {Wiegmann}},\ }\href {\doibase 10.1103/PhysRevB.88.241305} {\bibfield
  {journal} {\bibinfo  {journal} {Phys. Rev. B}\ }\textbf {\bibinfo {volume}
  {88}},\ \bibinfo {pages} {241305} (\bibinfo {year} {2013}{\natexlab{b}})},\
  \bibinfo {note} {\href{http://arxiv.org/abs/1309.5992}{\tt
  arXiv:1309.5992v1}}\BibitemShut {NoStop}%
\bibitem [{\citenamefont {Laughlin}(1981)}]{Laughlin1981}%
  \BibitemOpen
  \bibfield  {author} {\bibinfo {author} {\bibfnamefont {R.~B.}\ \bibnamefont
  {Laughlin}},\ }\href {\doibase 10.1103/PhysRevB.23.5632} {\bibfield
  {journal} {\bibinfo  {journal} {Phys. Rev. B}\ }\textbf {\bibinfo {volume}
  {23}},\ \bibinfo {pages} {5632} (\bibinfo {year} {1981})}\BibitemShut
  {NoStop}%
\bibitem [{\citenamefont {{Zabrodin}}\ and\ \citenamefont
  {{Wiegmann}}(2006)}]{Zabrodin2006}%
  \BibitemOpen
  \bibfield  {author} {\bibinfo {author} {\bibfnamefont {A.}~\bibnamefont
  {{Zabrodin}}}\ and\ \bibinfo {author} {\bibfnamefont {P.}~\bibnamefont
  {{Wiegmann}}},\ }\href {\doibase 10.1088/0305-4470/39/28/S10} {\bibfield
  {journal} {\bibinfo  {journal} {J. Phys. A: Math. Gen.}\ }\textbf {\bibinfo
  {volume} {39}},\ \bibinfo {pages} {8933} (\bibinfo {year} {2006})},\ \bibinfo
  {note} {\href{http://arxiv.org/abs/hep-th/0601009}{\tt arXiv:0601009v3}},\
  \Eprint {http://arxiv.org/abs/hep-th/0601009} {hep-th/0601009} \BibitemShut
  {NoStop}%
\bibitem [{\citenamefont {Furtado}\ \emph {et~al.}(1994)\citenamefont
  {Furtado}, \citenamefont {da~Cunha}, \citenamefont {Moraes}, \citenamefont
  {de~Mello},\ and\ \citenamefont {Bezzerra}}]{Furtado}%
  \BibitemOpen
  \bibfield  {author} {\bibinfo {author} {\bibfnamefont {C.}~\bibnamefont
  {Furtado}}, \bibinfo {author} {\bibfnamefont {B.~G.}\ \bibnamefont
  {da~Cunha}}, \bibinfo {author} {\bibfnamefont {F.}~\bibnamefont {Moraes}},
  \bibinfo {author} {\bibfnamefont {E.}~\bibnamefont {de~Mello}}, \ and\
  \bibinfo {author} {\bibfnamefont {V.}~\bibnamefont {Bezzerra}},\ }\href
  {\doibase http://dx.doi.org/10.1016/0375-9601(94)90432-4} {\bibfield
  {journal} {\bibinfo  {journal} {Phys. Lett. A}\ }\textbf {\bibinfo {volume}
  {195}},\ \bibinfo {pages} {90 } (\bibinfo {year} {1994})}\BibitemShut
  {NoStop}%
\bibitem [{\citenamefont {{Marques}}\ \emph {et~al.}(2001)\citenamefont
  {{Marques}}, \citenamefont {{Furtado}}, \citenamefont {{Bezerra}},\ and\
  \citenamefont {{Moraes}}}]{Furtado2}%
  \BibitemOpen
  \bibfield  {author} {\bibinfo {author} {\bibfnamefont {G.~d.~A.}\
  \bibnamefont {{Marques}}}, \bibinfo {author} {\bibfnamefont {C.}~\bibnamefont
  {{Furtado}}}, \bibinfo {author} {\bibfnamefont {V.~B.}\ \bibnamefont
  {{Bezerra}}}, \ and\ \bibinfo {author} {\bibfnamefont {F.}~\bibnamefont
  {{Moraes}}},\ }\href {\doibase 10.1088/0305-4470/34/30/306} {\bibfield
  {journal} {\bibinfo  {journal} {J. Phys. A: Math. Gen.}\ }\textbf {\bibinfo
  {volume} {34}},\ \bibinfo {pages} {5945} (\bibinfo {year} {2001})},\ \bibinfo
  {note} {\href{http://arxiv.org/abs/quant-ph/0012146}{\tt
  arXiv:quant-ph/0012146v1}},\ \Eprint {http://arxiv.org/abs/quant-ph/0012146}
  {quant-ph/0012146} \BibitemShut {NoStop}%
\bibitem [{\citenamefont {{Poux}}\ \emph {et~al.}(2014)\citenamefont {{Poux}},
  \citenamefont {{Ara{\'u}jo}}, \citenamefont {{Filgueiras}},\ and\
  \citenamefont {{Moraes}}}]{Poux}%
  \BibitemOpen
  \bibfield  {author} {\bibinfo {author} {\bibfnamefont {A.}~\bibnamefont
  {{Poux}}}, \bibinfo {author} {\bibfnamefont {L.~R.~S.}\ \bibnamefont
  {{Ara{\'u}jo}}}, \bibinfo {author} {\bibfnamefont {C.}~\bibnamefont
  {{Filgueiras}}}, \ and\ \bibinfo {author} {\bibfnamefont {F.}~\bibnamefont
  {{Moraes}}},\ }\href {\doibase 10.1140/epjp/i2014-14100-9} {\bibfield
  {journal} {\bibinfo  {journal} {Eur. Phys. J. Plus}\ }\textbf {\bibinfo
  {volume} {129}},\ \bibinfo {eid} {100} (\bibinfo {year} {2014})},\ \bibinfo
  {note} {\href{http://arxiv.org/abs/1405.6599}{\tt arXiv:1405.6599}},\ \Eprint
  {http://arxiv.org/abs/1405.6599} {arXiv:1405.6599 [cond-mat.mes-{H}all]}
  \BibitemShut {NoStop}%
\bibitem [{not()}]{note1}%
  \BibitemOpen
  \href@noop {} {}\bibinfo {note} {In a general orbifold setting, where $\gamma
  = m/n$ is rational, and $q \not =0$, the density is a finite sum $\rho = \rho
  _\infty e^{-x} \sum_{k=0} ^{n-1} e^{ x \omega _n^{k} }\sum_{k' = 0}^{m-1}
  \omega _n^{ - k (k' + q)/\gamma } \protect \frac {\gamma \left ((k' +
  q)/\gamma , x \omega _n^{k}\right )}{\Gamma \left ( (k' + q)/\gamma \right )}
  $ where $\gamma (s, x) = \intop_{0}^{x} t^{s - 1}e^{ - t} dt$ is the lower
  incomplete gamma function.}\BibitemShut {Stop}%
\end{thebibliography}%

\end{document}